%% file: main.tex
\documentclass[a4paper,11pt]{article}
\pdfoutput=1 
\usepackage{jheppub} 
\usepackage{makecell}   
\usepackage{booktabs}
\usepackage{lscape}
\usepackage[capitalise]{cleveref}
\usepackage[numbers]{natbib}
\usepackage[T1]{fontenc} 
\usepackage{subcaption}
\usepackage{siunitx}
\usepackage{bbold}
\usepackage{float}
\usepackage{feynman}
\usepackage{xcolor}
\usepackage{adjustbox}
\usepackage{blindtext}
\usepackage{multirow}
\usepackage{multicol}
\usepackage{verbatim}
\usepackage{amsmath,mathtools}
\usepackage{relsize}
\usepackage{bm, nicefrac}
\usepackage{hyperref}
\usepackage{cleveref}
\usepackage[normalem]{ulem}
\usepackage{subcaption}
\usepackage{lipsum}
\usepackage{comment}
\usepackage{tabularx}
\usepackage[compat=1.1.0]{tikz-feynman}

\renewcommand{\phi}{\ensuremath{\varphi}}
\newcommand{\sss}{\scriptscriptstyle}

\newcommand{\QQ}{\ensuremath{\mathcal{Q}}}

\newcommand{\Qpp}[2]{\QQ_{\sss #1}^{\sss #2}}

\newcommand{\expnumber}[2]{{#1}\mathrm{e}{#2}}

\pgfmathsetmacro\sizedot{1.1}
\pgfmathsetmacro\sizesqdot{1.5}
\pgfmathsetmacro\sizecrodot{1.0}

\newcommand{\order}[1] {\mathcal{O }\left( #1 \right)}

\title{Constraining four-heavy-quark operators with top-quark, Higgs, and electroweak precision data}

\author[a]{Stefano Di Noi,}
\author[b]{Hesham El Faham,}
\author[c]{Ramona Gr\"ober,}
\author[d,e]{Marco Vitti, }
\author[b]{Eleni Vryonidou}

\affiliation[a]{Institute for Theoretical Physics, Karlsruhe Institute of Technology (KIT),\\
Wolfgang-Gaede Stra\ss{}e 1, Karlsruhe D-76131, Germany}
\affiliation[b]{Department of Physics and Astronomy, University of Manchester,\\
Oxford Road, Manchester M13~9PL, United Kingdom}
\affiliation[c]{Dipartimento di Fisica e Astronomia “G. Galilei”, Universit\`a di Padova, and Istituto Nazionale di Fisica Nucleare, Sezione di Padova,\\
Via Marzolo 8, I-35131 Padova, Italy}
\affiliation[d]{Institute for Theoretical Particle Physics, Karlsruhe Institute of Technology (KIT), \\
Wolfgang-Gaede Stra\ss{}e 1, Karlsruhe D-76131, Germany}
\affiliation[e]{Institute for Astroparticle Physics, Karlsruhe Institute of Technology (KIT),\\
 Hermann-von-Helmholtz-Platz 1 Eggenstein-Leopoldshafen, Karlsruhe D-76344, Germany}

\emailAdd{stefano.dinoi@kit.edu}
\emailAdd{hesham.elfaham@manchester.ac.uk}
\emailAdd{ramona.groeber@pd.infn.it}
\emailAdd{marco.vitti@kit.edu}
\emailAdd{eleni.vryonidou@manchester.ac.uk}

\preprint{
\begin{flushright}
KA-TP-12-2025, 
TTP25-021,
P3H-25-048 
\end{flushright}
}

\abstract{We establish constraints on the dimension-six four-heavy-quark operators in the Standard Model Effective Field Theory (SMEFT) by synthesising LHC measurements of top-quark and single-Higgs production with electroweak precision observables. 
We scrutinise the choice of the \(\gamma_5\) scheme in single-Higgs calculations, demonstrating its non-negligible impact on SMEFT fits.} 

\keywords{SMEFT, higher-loop computations, Monte Carlo, collider physics}

\begin{document}

\maketitle
\flushbottom

\section{Introduction}
\label{sec:intro} 
The absence of direct evidence for new light particles beyond the Standard Model (SM) at the Large Hadron Collider (LHC) has motivated a campaign of indirect searches in the SM Effective Field Theory (SMEFT) framework. In SMEFT, new-physics effects are parametrised through a series of higher-dimensional operators modifying the interactions of the SM particles. Thus, SMEFT provides a systematic and model-agnostic way of probing new physics in the absence of new light states. A key strength of the SMEFT framework lies in its ability to correlate effects across different sectors of the SM interactions. In order to fully exploit its potential in identifying signatures of physics beyond the SM, global analyses and interpretations within the SMEFT paradigm have become essential, prompting significant ongoing efforts~\cite{deBlas:2016ojx, Ellis:2020unq, Brivio:2021alv, Ethier:2021bye, Celada:2024mcf,Bartocci:2024fmm,terHoeve:2025gey,deBlas:2025xhe}. 

Such global interpretations are crucial, as they can reveal potential signs of new physics or, at the very least, place constraints on the energy scale at which new physics could appear by setting bounds on the Wilson Coefficients (WCs) of higher-dimensional operators. Additionally, global interpretations help identify sectors with greater potential for deviations from the SM by highlighting the least constrained operator classes. Notably, operators involving four heavy-quark fields stand out as among the least constrained by current experimental data. These induce contact interactions of four top quarks ($t\bar{t}t\bar{t}$), four bottom quarks ($b\bar{b}b\bar{b}$), and interactions involving a top-quark pair with a bottom-quark pair ($t\bar{t}b\bar{b}$).

Results from global fits indicate that the new-physics scale associated with this class of operators can be as low as a few hundred GeV; see, for example, the recent global analysis by the SMEFiT collaboration~\cite{Celada:2024mcf}. These loose constraints arise because, at tree level, these interactions are predominantly probed by $t\bar t t\bar t$ or $t\bar t b\bar b$ production, which suffer from large experimental uncertainties~\cite{ATLAS:2023ajo,CMS:2023ftu,CMS:2019eih}. Moreover, these inclusive measurements are not sufficient to distinguish different colour and chirality structures in contact interactions, leading to flat directions that weaken the constraints.

These findings–together with model-building arguments suggesting that new physics might be top-philic~\cite{Lillie:2007hd,Kumar:2009vs,Banelli:2020iau,Darme:2021gtt,Chung:2023gcm,Choudhury:2024mox}–have motivated indirect probes of the four-heavy contact operators via their higher-loop contributions to observables that are measured more precisely than multi-top-quark production. In particular, the effects of four-heavy-quark operators on electroweak precision observables (EWPO)~\cite{deBlas:2015aea,Dawson:2019clf,Haisch:2024wnw}, single-Higgs production in gluon fusion~\cite{Alasfar:2022zyr,DiNoi:2023ygk}, top-quark-pair production~\cite{Degrande:2020evl,Degrande:2024mbg} and flavour observables~\cite{Silvestrini:2018dos,Haisch:2024wnw} have been computed. These studies demonstrate that such indirect probes supply information complementary to four-top-quark production and must be included to obtain tighter limits on the strength of these interactions.

In this work we carry out a fit that combines direct and indirect probes, thereby exploiting their complementarity. Specifically, we include leading-order (LO) contributions to $t\bar t t\bar t$ and $t\bar t b\bar b$ production; next-to-leading-order (NLO) contributions to $t\bar t$ and $t\bar t H$ production; two-loop effects in gluon-fusion Higgs production ($gg\!\to\! H$) and Higgs decays, together with the two-loop contributions to EWPO.

The chiral nature of the four-fermion contact interactions in the SMEFT demands particular care in loop computations, because the Dirac algebra necessarily involves $\gamma_5$.
Since $\gamma_5$ is intrinsically four-dimensional, one must define a consistent continuation to $d=4-2\epsilon$ dimensions.  
In this work, we adopt two distinct continuation schemes: the \emph{na\"ive dimensional regularisation} (NDR) scheme~\cite{CHANOWITZ1979225} and the \emph{Breitenlohner–Maison–’t Hooft–Veltman} (BMHV) scheme~\cite{THOOFT1972189,Breitenlohner:1977hr}.

Provided that each scheme is implemented consistently, any differences in EFT matrix elements can be traced either to the scheme-dependent definition of the WCs~\cite{DiNoi:2023ygk} or to finite terms specific to the chosen renormalisation prescription~\cite{DiNoi:2025uan}. This is due to the fact that NDR and BMHV predictions can be different in loop computations. We demonstrate this explicitly for the two-loop process of single-Higgs production and decay mediated by four-heavy-quark operators: in a dedicated fit, the bounds extracted for the WCs differ between the NDR and BMHV schemes. Such scheme dependence has been studied also in the context of di-Higgs production at the LHC~\cite{Heinrich:2023rsd}, flavour physics~\cite{Buras:1991jm,Buras:1992tc,Ciuchini:1993ks,Adel:1993ah,Ciuchini:1993fk}.
It is worth noting that the choice of the $\gamma_{5}$ continuation scheme is not the only source of scheme dependence in loop computations~\cite{Corbett:2021cil,Martin:2023fad,Biekotter:2023xle}.

The paper is organised as follows: in~\cref{sec:intro_smeft}, we state our flavour assumptions, introduce the SMEFT operators relevant to this study, and outline the computational setup. \Cref{sec:aspect_of_processes} presents our predictions for the processes under consideration and summarises the analytic expressions for the four–top-quark operators that modify the $ggH$ and $\gamma\gamma H$ couplings in both the NDR and BMHV \(\gamma_{5}\) schemes. The scheme dependence of the resulting bounds on the WCs is examined in~\cref{sec:interpretation}. Our fitting method and core results are detailed in~\cref{sec:fit}. Finally,~\cref{sec:conclusions} summarises our findings.

\section{Theoretical framework and computation setup} 
\label{sec:intro_smeft}
In this section, we discuss the SMEFT theoretical employed, along with the technical details underlying our computations.
\subsection{SMEFT framework}
A generic SMEFT Lagrangian, including terms up to $\mathcal{O}(\Lambda^{-4})$, can be written as
\begin{equation}
\mathcal{L}_{\rm SMEFT}
= \mathcal{L}_{\rm SM}
+ \sum_i \frac{c_i^{(6)}\,\mathcal{O}_i^{(6)}}{\Lambda^2}
+ \sum_j \frac{c_j^{(8)}\,\mathcal{O}_j^{(8)}}{\Lambda^4}
+ \mathcal{O}\bigl(\Lambda^{-6}\bigr)\,,
\label{eq:EFT_Lagrangian}
\end{equation}
where \(c_i^{(D)}\) and \(\mathcal{O}_i^{(D)}\) are the WCs and SMEFT operators of mass dimension \(D\), respectively, and \(\Lambda\) denotes the scale of new physics. Restricting to the dimension-six operators, the SMEFT prediction for cross section can be parametrised as
\begin{equation}
  \sigma_{\mathrm{SMEFT}} \;=\; \sigma_{\mathrm{SM}} \;+\; \sigma_{\mathrm{int}}^{(i)}\,\frac{c_i}{\Lambda^2}
    \;+\;\sigma_{\mathrm{quad}}^{(i)}\,\frac{c_i^2}{\Lambda^4}
    \;+\;\sigma_{\mathrm{cross}}^{(i,j)}\, \frac{c_i\,c_j}{\Lambda^4} \,,
\label{eq:theory_prediction}
\end{equation}
Here \(\sigma_{\mathrm{int}}\) originates from the interference between the SM and dimension-six SMEFT amplitudes scaling as \(\Lambda^{-2}\), while \(\sigma_{\mathrm{quad}}\) and \(\sigma_{\mathrm{cross}}\) denote the diagonal (\(c_i^2\)) and off-diagonal (\(c_i c_j\)) quadratic contributions scaling as \(\Lambda^{-4}\). An analogous parametrisation will be adopted for the partial widths, $\Gamma$. In all our results we set $\Lambda=1$ TeV.

We use a specific flavour assumption of the SMEFT focused on top-quark interactions:
\begin{equation}
    \begin{aligned}
    U(3)_l\times U(3)_e \times U(2)_q\times U(2)_u\times U(3)_d \equiv U(2)^{2}\times U(3)^{3},
    \end{aligned}
\label{eq:flav_symmetry}
\end{equation}
where the subscripts denote the five-fermion representations of the SM. This minimal relaxation of the $U(3)^5$ group allows for top-quark-chirality-flipping interactions, such as dipole interactions and modifications to the top-Yukawa coupling. In the widely-used Warsaw basis~\cite{Grzadkowski:2010es}, the four-heavy subclass of dimension-six four-fermion operators are defined as follows:
\begin{equation}
\label{eq:dim64f_op_warsaw}
\begin{aligned}
    \Qpp{qq}{1(ijkl)} &= (\bar{q}_i \gamma^{\mu} q_{j})(\bar{q}_k \gamma_{\mu} q_{l}), & 
    \Qpp{qq}{3(ijkl)} &= (\bar{q}_i \gamma^{\mu} \tau^{I} q_{j})(\bar{q}_k \gamma_{\mu} \tau^{I} q_{l}), \notag \\
    \Qpp{qu}{1(ijkl)} &= (\bar{q}_i \gamma^{\mu} q_{j})(\bar{u}_k \gamma_{\mu} u_{l}), & 
    \Qpp{qu}{8(ijkl)} &= (\bar{q}_i \gamma^{\mu} T^{A} q_{j})(\bar{u}_k \gamma_{\mu} T^{A} u_{l}), \notag \\
    & & \Qpp{uu}{(ijkl)} &= (\bar{u}_i \gamma^{\mu} u_{j})(\bar{u}_k \gamma_{\mu} u_{l}),
\end{aligned}
\end{equation}
where only third-generation quark fields are understood and we have used the notation $\mathcal{Q}$ to denote operators written in said Warsaw basis, with the corresponding WCs denoted as $C_{i}$. However, in this work, we use operators aligned with the \texttt{dim6top}~\cite{Aguilar-Saavedra:2018ksv} and \texttt{SMEFTatNLO}~\cite{Degrande:2020evl} conventions, hereafter referred to as the `top basis'—in which the operators are written as $\mathcal{O}_i$ with Wilson coefficients $c_i$. The translations of four-heavy coefficients at tree level from the Warsaw basis to the top basis are as follows: 
\begin{equation}
    \begin{aligned}
    c^1_{QQ} &= 2 C^{(1)}_{qq} - \frac{2}{3} \, C^{(3)}_{qq}, & 
    c^1_{Qt} &= C^{(1)}_{qu}, \\
    c^1_{tt} &= C^{(1)}_{uu}, & 
    c^8_{QQ} &= 8 C^{(3)}_{qq}, \\
    c^8_{Qt} &= C^{(8)}_{qu}.
    \end{aligned}
\label{eq:dim64f_smeftatnlo_basis_main}
\end{equation}
The corresponding state-of-the-art constraints are reported in the recent global fit of Ref.~\cite{Celada:2024mcf}.
We note here that, in all our computations, we adopt the definition of the four-heavy operator, $\mathcal{O}_{QQ}^8$, in terms of Warsaw-basis operators, i.e.\ $\mathcal{O}_{QQ}^8=\mathcal{Q}_{qq}^{(3)}/8+\mathcal{Q}_{qq}^{(1)}/24$, rather than $\mathcal{O}_{QQ}^8=(1/2)(\bar{Q}\gamma^{\mu}T^{A}Q)(\bar{Q}\gamma_{\mu}T^{A}Q)$. The two expressions differ by an evanescent operator, as also discussed in Ref.~\cite{Haisch:2024wnw}. Numerical results can differ between the two definitions when the evanescent operator contributes. All our results are consistent with the former definition, and we will comment on this further.

\subsection{Computation setup}
\label{sec:computationalsetup}
For all our predictions, we utilise \texttt{MadGraph5\_aMC@NLO}~\cite{Alwall:2014hca} and the \texttt{SMEFT@NLO}~\cite{Degrande:2020evl} package, with the exceptions being $gg \to H$, for which we employ the analytic expression given in Ref.~\cite{DiNoi:2023ygk} and the EWPO, for which we use the expressions in~\cref{eq:ewpo-shifts} extracted from~\cite{Dawson:2019clf, Haisch:2024wnw} -- see dedicated discussions below. The parton distribution functions (PDF) set \texttt{NNPDF3.1} in the five-flavour scheme at NLO with $\alpha_{s}(m_{Z})=0.118$~\cite{Ball:2017nwa} is used as input for all Monte Carlo (MC) simulations through the \texttt{LHAPDF} interface~\cite{Buckley:2014ana}. All computations are carried out in the $G_F$ scheme~\cite{Denner:2000bj,Dittmaier:2001ay}, which is recommended for SMEFT analyses by the LHC EFT WG~\cite{Brivio:2021yjb}. The Fermi constant value is set to $G_{F}=1.16637 \times 10^{-5} \, \rm GeV^{-2}$. The masses of the Higgs boson and the top quark are set to $m_H=125 \, \text{GeV}$ and $m_t=172 \, \text{GeV}$, respectively. 

As will be relevant later, it is important to note that in \texttt{SMEFT@NLO}~\cite{Degrande:2020evl}, in $d\ne 4$ dimensions, $\gamma_5$ is treated as anti-commuting, and the cyclic property of Dirac matrices traces is not maintained following the KKS scheme~\cite{Korner:1991sx}. The latter is understood to be equivalent to the NDR scheme supplemented with a fixed reading point.

The factorisation and renormalisation scales, $\mu_F$ and $\mu_R$, are set to half of the sum of the masses of the final state particles. The scale $\mu_{\text{EFT}}$ introduced in the counterterms of the WCs is set to $\mu_R$ to ensure $\overline{\rm MS}$ renormalisation for the EFT poles~\cite{Degrande:2024mbg}. The effects of renormalisation group equations (RGEs) on the WCs are not considered in this work, and we interpret the WCs at a typical electroweak (EW) scale. We refer to Refs.~\cite{Aoude:2022aro,Maltoni:2024dpn,DiNoi:2023onw,DiNoi:2024ajj} for RGE effects in this context. 
Considering the EWPO at scale $Q \sim m_Z$ and four-top-quark production at $Q \sim 2 m_t$, we adopt scales within a factor of four of each other and thus expect RGE effects to be under control. 

\section{Studied processes}
\label{sec:aspect_of_processes}
In this section, we review the characteristic features of the LHC processes analysed in this work: four‐top‐quark production; top‐quark pair production; top‐quark pair production in association with a Higgs boson; single‐Higgs production with its subsequent decay. Finally, we discuss the EWPO predictions employed in our analysis. 

We keep our discussion of top-quark processes concise, since these computations are already well established and extensively discussed in the literature. For the EWPO predictions, however, we explicitly report the individual contributions to each of the WCs, as we are not aware of any previous work where this has been done. In~\cref{app:ewpo}, we provide the details of how these contributions were extracted using the EWPO two-loop results of Ref.~\cite{Haisch:2024wnw}. Finally, our treatment of single-Higgs production is comparatively more extensive, as we use it to illustrate the emergence of the $\gamma_{5}-$scheme dependence in the computation—an aspect that is central to the discussion in the following section. All results discussed in this section are presented in the NDR scheme, unless explicitly stated otherwise, as in the context of single-Higgs production.

\subsection{Four-top-quark production}
\label{sec:tttt_comp}
Our $pp \to t\bar{t}t\bar{t}$ predictions are listed in~\cref{tab:all_pred_incl} of~\cref{app:numerical_data}.  
As demonstrated in Ref.~\cite{Aoude:2022deh}, subleading terms stemming from the interference of four‑fermion operators with weakly mediated SM amplitudes play a non‑negligible role; all such contributions are therefore included in our calculation. It is worth noting that, unlike $t\bar{t}$ production, the richer colour structure of the $pp \to t\bar t t\bar t$ process allows colour‑singlet operators to interfere with the QCD SM amplitudes already at LO.  

In the pure SM, subleading EW contributions almost exactly cancel among themselves~\cite{Frederix:2017wme} making the leading NLO QCD prediction highly reliable.  
In the EFT, however, such cancellation is not guaranteed because the SMEFT operators alter the kinematic structure of the amplitudes.  
A fully consistent NLO prediction in the SMEFT would thus require the complete set of NLO QCD \emph{and} EW corrections—an undertaking that is presently beyond the reach of existing automated tools. Therefore, we employ only tree-level predictions for this process. 

\subsection{Top-quark pair production (and in association with a Higgs)}
\label{sec:tt_comp}
In $pp \to t\bar{t}(H)$, four‑fermion operators that involve a third‑generation doublet contribute at tree level through $b$‑quark–initiated amplitudes. The interference of colour‑singlet four-fermion operators with the SM vanishes at tree level when only purely QCD‑induced amplitudes are considered~\cite{Brivio:2019ius}, because the top quarks are always produced in a colour‑octet configuration.\footnote{For weak‑mediated Born amplitudes, i.e.\ $t\bar{t}$ production via an $s$‑channel weak boson, colour‑singlets can already interfere at LO. These contributions are generally expected to be subdominant and are not considered here.}   
At NLO in QCD, however, both real and virtual corrections alter the colour flow and can induce a non‑zero interference for colour‑singlets.

Inclusive predictions for all processes as well as the differential predictions in the Higgs boson transverse momenta, $p_{T}^{H}$, for both the SM and the SMEFT are collected in~\cref{tab:all_pred_incl,tab:tth_pth_pred} of~\cref{app:numerical_data}.\footnote{Differential predictions in the top-quark-pair invariant mass, \(m_{t\bar{t}}\), for \(t\bar{t}\) production are omitted here for brevity, but can be provided upon request.} The SMEFT results are separated into linear, $\mathcal{O}(\Lambda^{-2})$, and (off-) diagonal quadratic, $\mathcal{O}(\Lambda^{-4})$, terms.
As previously mentioned, at LO, the operators under study contribute only through $b$‑quark–initiated channels so the diagonal‑quadratic piece from $c_{tt}^{1}$ vanishes.  
Because $c_{tt}^{1}$ first appears at one loop, obtaining its quadratic contribution would require squaring the loop amplitudes rendering it beyond the perturbative order considered in this work.

We observe that $c_{Qt}^{1}$ dominates the SMEFT corrections at NLO, providing by far the largest linear contribution to the cross section. This is the case in both $t\bar{t}$ and $t\bar{t}H$ production.  
The dominance of this contribution can be traced back to cancellations among different partonic channels and phase-space regions for all coefficients except $c_{Qt}^{1}$~\cite{Degrande:2020evl}.
Moreover, the contribution of the right‑handed four‑top-quark operator, $\mathcal{O}_{tt}^{1}$, features some strong cancellation between the gluon‑ and quark‑initiated channels in $t\bar{t}H$ production.  
This cancellation amplifies the scale dependence and results in sizeable QCD uncertainties for the linear contribution, as illustrated in~\cref{tab:all_pred_incl}.  
The same pattern is visible differentially in one of the $p_{T}^{H}$ bins in~\cref{tab:tth_pth_pred}.  
In $t\bar{t}$ production, this effect is absent in the inclusive rate.
Moving to the off-diagonal quadratic terms, at LO, interference between colour‑singlet and colour‑octet structures—whether between SM and EFT amplitudes or between two different EFT operators—vanishes exactly, and so only singlet–singlet and octet–octet combinations survive, as shown in~\cref{tab:all_pred_incl}.  
At NLO, real emissions or virtual gluon exchange can mix the colour flows, generating singlet–octet cross terms which are numerically tiny–cross terms consistent with zero at the $2\sigma$ level are therefore omitted from the tables for clarity.

\begin{figure}[ht]
    \centering
    \includegraphics[width=0.40\linewidth]{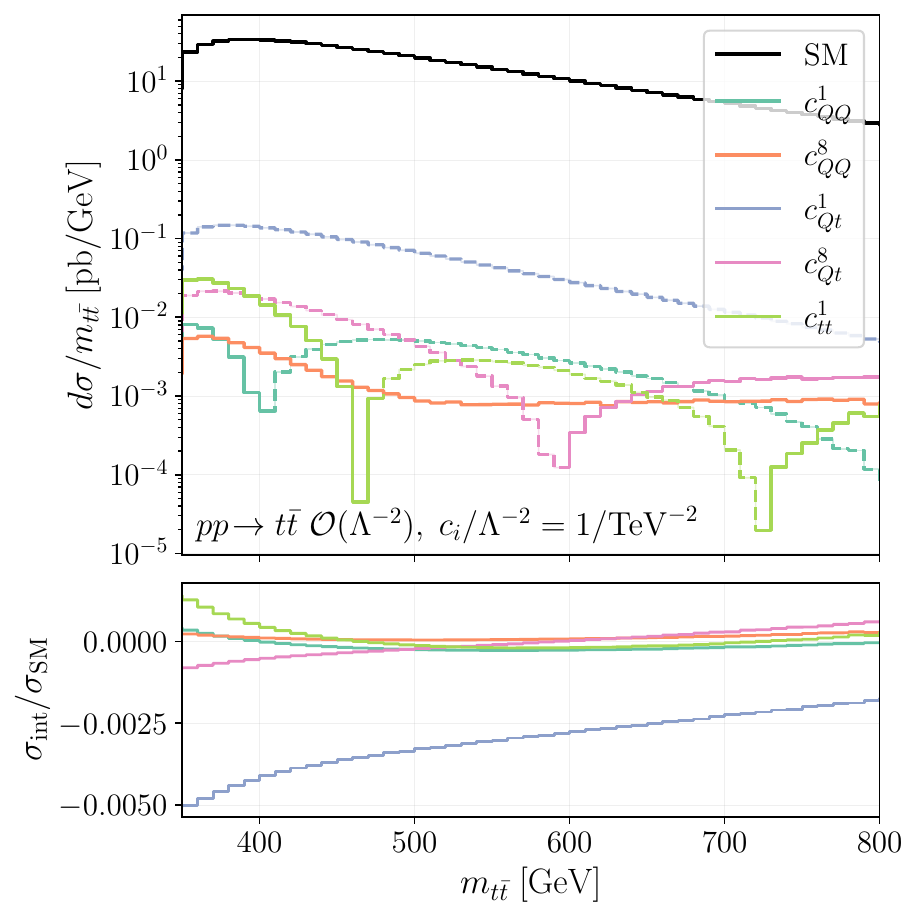}
    \includegraphics[width=0.40\linewidth]{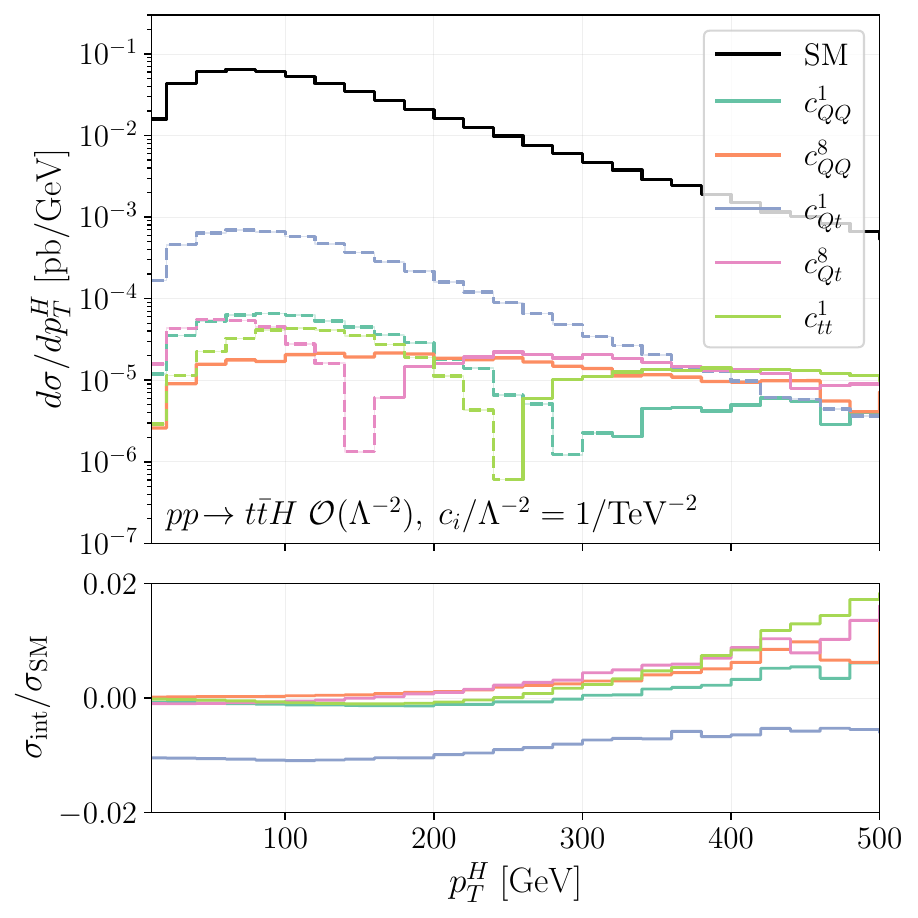}
    \caption{Differential distributions of the top-quark-pair invariant mass in $t\bar{t}$ (left) and of the Higgs transverse momentum in $t\bar{t}H$ (right). The curves show the linear SMEFT contributions of the five four-heavy operators, compared to the NLO SM prediction. The absolute values are plotted, and the dashed lines indicate where the interference becomes destructive—that is, where the contributions are negative.}
    \label{fig:diff_mtt_pth}
\end{figure}

Finally, in~\cref{fig:diff_mtt_pth}, the differential distributions of the top–quark-pair invariant mass, $m_{t\bar{t}}$, and the Higgs transverse momentum, $p_T^{H}$, are displayed in the left and right panels, respectively. 
For ${\cal O}(1)$ WCs, the four–heavy operators modify the SM prediction by at most the percent level.  This mild impact is expected, given both the one-loop suppression and the small $b$-quark parton density driving these contributions. The dominant correction comes from $\mathcal{O}_{Qt}^{1}$: its contribution is significant in the low-invariant-mass region of the $t\bar t$ spectrum while it induces an essentially flat shift in the $p_T^{H}$ distribution.  All other operators yield sub-percent effects; in some kinematic bins their contributions even change sign, as indicated by the dashed curves in the figures.

\subsection{Electroweak precision observables} 
\label{sec:EWPO} 
Our EWPO predictions are obtained at the two‐loop level leveraging the work of Ref.~\cite{Haisch:2024wnw} and reported here in~\cref{eq:ewpo-shifts}. These results are quoted in the NDR scheme. It is also important to emphasise that all EWPO results were extracted in the original Warsaw basis from Ref.~\cite{Haisch:2024wnw}. These results are expected to be consistent with our \texttt{SMEFT@NLO} calculations. Although the latter employs the rotations defined in~\cref{eq:dim64f_smeftatnlo_basis_main}, it ultimately remains formulated in the Warsaw basis. In summary, extracting the EWPO results directly in either the original Warsaw basis or the top basis is not expected to affect the WCs contributions, with the sole exception of the four-heavy colour-octet operator. The original Warsaw basis choice for said operator will differ by an evanescent contribution compared to its contribution defined with two colour-octet currents–see discussion below~\cref{eq:dim64f_smeftatnlo_basis_main}.
The reason is due to the mapping in~\cref{eq:dim64f_op_warsaw} being a tree-level one (cf.\ Ref.~\cite{Haisch:2024wnw}). 
We further note that this evanescent term can be numerically significant, and a substantial shift in the corresponding
contribution would be expected had the results in the top basis conventions of Ref.~\cite{Haisch:2024wnw} been employed from the outset.
\begin{equation}
\label{eq:ewpo-shifts}
\begin{aligned}
\delta\Gamma_Z^{b\bar b}
&= 9.5412\times10^{-4}\,c_{QQ}^{1}
  +1.0098\times10^{-4} \,c_{QQ}^{8}
  -1.1409\times10^{-3}\,c_{Qt}^{1}\\
&\quad
  +4.4956\times10^{-7}\,c_{Qt}^{8}
  -3.12\times10^{-6}\,c_{tt}^{1},
\\[6pt]
\delta R_c
&=-9.699\times10^{-5}\,c_{QQ}^{1}
  -1.0265\times10^{-5}\,c_{QQ}^{8}
  +1.1598\times10^{-4}\,c_{Qt}^{1}\\
&\quad
  -4.5701\times10^{-8}\,c_{Qt}^{8}
  +3.1718\times10^{-7}\,c_{tt}^{1},
\\[6pt]
\delta R_l
&= 1.1688\times10^{-2}\,c_{QQ}^{1}
  +1.2371\times10^{-3} \,c_{QQ}^{8}
  -1.3977\times10^{-2}\,c_{Qt}^{1}\\
&\quad
  +5.5074\times10^{-6}\,c_{Qt}^{8}
  -3.8222\times10^{-5}\,c_{tt}^{1},
\\[6pt]
\delta R_b
&= 4.4158\times10^{-4}\,c_{QQ}^{1}
  +4.6736\times10^{-5} \,c_{QQ}^{8}
  -5.2803\times10^{-4}\,c_{Qt}^{1}\\
&\quad
  +2.0806\times10^{-7}\,c_{Qt}^{8}
  -1.444\times10^{-6}\,c_{tt}^{1},
\\[6pt]
\delta A_b
&= 2.4597\times10^{-4}\,c_{QQ}^{1}
  +3.2227\times10^{-5} \,c_{QQ}^{8}
  -2.9442\times10^{-4}\,c_{Qt}^{1}\\
&\quad
  +5.834\times10^{-7}\,c_{Qt}^{8}
  +5.7326\times10^{-5}\,c_{tt}^{1},
\\[6pt]
\delta A_{b,\mathrm{FB}}
&= 2.5306\times10^{-4}\,c_{QQ}^{1}
  +8.3078 \times10^{-5} \,c_{QQ}^{8}
  -3.0434\times10^{-4}\,c_{Qt}^{1}\\
&\quad
  +5.3495\times10^{-6}\,c_{Qt}^{8}
  +5.2565\times10^{-4}\,c_{tt}^{1}\,.
\end{aligned}
\end{equation}

The definitions of the observables, as well as details of the computation and numerical inputs, are provided in~\cref{app:ewpo}. We note that (off-)diagonal quadratic EFT contributions are strongly suppressed relative to the linear contributions shown in~\cref{eq:ewpo-shifts}.
Moreover, the contributions of $c_{Qt}^{8}$ and $c_{tt}^{1}$ are purely two-loop induced, as they do not contribute to the EWPO at one loop. 

Although the one‐loop contributions are not shown here–see~\cref{app:ewpo} for said contributions–in comparison, we find the two‐loop corrections to be negligible for all observables except for the bottom‐quark asymmetry, \(A_b\), and its forward–backward counterpart, \(A_{b,\mathrm{FB}}\), where they are significant.

\subsection{Single-Higgs production in gluon-fusion and Higgs decays}
\label{sec:h_comp}
\begin{figure}[ht]
\centering
    \begin{subfigure}{0.3\textwidth}
        \centering
        \input{Figures/Fig2.1}
        \caption{Higgs-top-quark coupling} \label{fig:4t_2loop_yt}
    \end{subfigure}
\begin{subfigure}{0.3\textwidth}
\centering
\input{Figures/Fig2.2}
        \caption{Top-quark propagator} \label{fig:4t_2loop_mt}
    \end{subfigure}
    \begin{subfigure}{0.3\textwidth}
        \centering
        \input{Figures/Fig2.3}
        \caption{Gluon-top-quark vertex}\label{fig:4t_2loop_gtt}
    \end{subfigure}
    \caption{Feynman diagrams illustrating the corrections to $gg\!\to\! H$ induced by four-top-quark SMEFT operator insertions.} \label{fig:4t_2loop}
\end{figure}
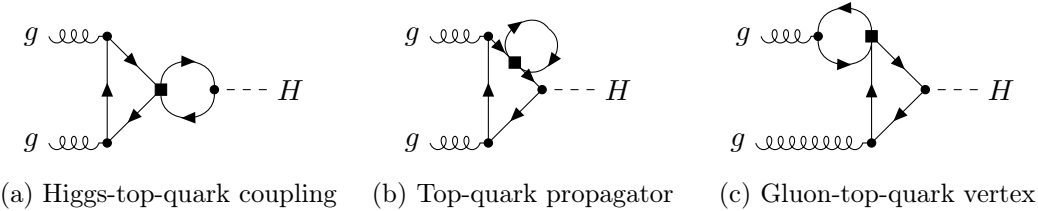

Higgs production via gluon fusion and its decays into gluons and photons are loop–induced already at LO. Four–top-quark operators contribute for the first time through two-loop diagrams, such as those shown in~\cref{fig:4t_2loop} for the production process. 

For single-Higgs production, we adopt the results of Ref.~\cite{DiNoi:2023ygk}. A key challenge in these computations arise from the presence of $\gamma_5$ in the loop amplitudes rendering a delicate treatment necessary when dimensional regularisation is used.
It has been shown in Ref.~\cite{DiNoi:2023ygk} that isolated contributions of four-top-quark operators depend on the continuation scheme of the $\gamma_5$ matrix to $d = 4-2\epsilon$ dimensions. The reference studied two schemes: the NDR~\cite{CHANOWITZ1979225} and the BMHV scheme~\cite{THOOFT1972189, Breitenlohner:1977hr}. Whilst the former is algebraically inconsistent in the presence of traces involving six or more $\gamma^\mu$ 
matrices~\cite{Korner:1991sx,Kreimer:1993bh,Chen:2023ulo,Chen:2024zju}, the latter remains consistent but the regulator spuriously breaks chiral symmetries and hence requires symmetry-restoring counterterms~\cite{Belusca-Maito:2020ala, Belusca-Maito:2021lnk, Cornella:2022hkc, Belusca-Maito:2023wah,OlgosoRuiz:2024dzq,vonManteuffel:2025swv,Ebert:2024xpy}. 

The continuation-scheme dependence is expected to cancel upon matching, once a process-specific set of operators at a given loop order is included.\footnote{This applies to scenarios in which UV divergences are absent in the UV model, a feature that is guaranteed for the operators considered here by the superficial degree of divergence~\cite{DiNoi:2025uan}.}
We adopt the loop-order definitions of Refs.~\cite{Arzt:1994gp, Buchalla:2022vjp}, which requires the assumption of weakly interacting and renormalisable UV models. 
Scheme-independent results can be achieved by including operators that enter at a lower loop order. In~\cref{app:matrixelements}, we specify those additional operators for the case of $ggH$ and $\gamma \gamma H$ and we provide results for the complete contribution to the renormalised matrix element, together with auxiliary quantities not defined in this section. Since the focus of the present paper is on the four-top operators, in this section we only show the contribution of the latter to the renormalised matrix element, which we denote as $\mathcal{M}^{ggH,\text{4t}}_{\text{OS}}$ and it reads as follows:
\begin{equation}
\label{eq:MgghOS4t}
    \mathcal{M}^{ggH,\text{4t}}_{\text{OS}} =    
    \left( c_{Qt}^{1} + \left(c_F-\dfrac{c_A}{2} \right) c_{Qt}^{8}\right)
    \mathcal{K}_{tG}   \frac{1}{\Lambda^2} \mathcal{M}_{tG}^{ggH} + \left( c_{Qt}^{1} + c_F c_{Qt}^{8}\right) \frac{1}{\Lambda^2} \left( B_{ggH}+ \mathcal{K}_{t \phi} \right)\mathcal{M}^{ggH}_{\mathrm{SM}} ,
\end{equation}
\begin{equation}
    \mathcal{K}_{tG} =
    \begin{cases}
    \frac{\sqrt{2} m_t g_s}{16 \pi^2
    v} & \text{(NDR)}\\
    0 & \text{(BMHV)},
    \end{cases} \qquad 
    \mathcal{K}_{t\phi} =
    \begin{cases}
    \frac{m_H^2-4 m_t^2}{16 \pi ^2} & \text{(NDR)}\\
    0 & \text{(BMHV)},
    \end{cases} 
    \label{eq:Kggh}
\end{equation}
\begin{equation} 
\label{eq:F4t}
    B_{ggH}= \frac{4m_t^2 - m_H^2}{8 \pi^2} \left( \beta \log \left( \frac{\beta - 1 }{\beta + 1} \right)  + 2 +  \log \left(\frac{{\mu}_{R}^2}{m_t^2} \right) \right), \quad  \beta= \sqrt{1- \frac{4 m_t^2}{m_H^2}}. 
\end{equation}
The matrix element in~\cref{eq:MgghOS4t} has been obtained in the on-shell (OS) renormalisation scheme for the top-quark mass, whilst the operators are renormalised in the $\overline{\mathrm{MS}}$ scheme. 
We note that this differs from the renormalisation scheme of Ref.~\cite{DiNoi:2023ygk}. The scheme dependence in~\cref{eq:MgghOS4t} and in the following~\cref{eq:MgammagammahOS4t} is parametrised by the $\mathcal{K}$ terms in~\cref{eq:Kggh}. 
The scheme dependence arises, in principle, from all the diagrams sketched in Fig.~\ref{fig:4t_2loop}. In particular, the diagram topology in Fig.~\ref{fig:4t_2loop_yt} gives the scheme-dependent contribution associated to $\mathcal{K}_{t\phi}$, the diagram topology in Fig.~\ref{fig:4t_2loop_mt} is nullified by the on-shell top mass counterterm and the diagram topology in Fig.~\ref{fig:4t_2loop_gtt} gives the scheme-dependent contribution associated to $\mathcal{K}_{tG}$. A detailed diagram-by-diagram analysis can be found in Ref.~\cite{DiNoi:2023ygk}, to which we refer the interested reader.

Finally, we note that only the two operators $\mathcal{O}_{Qt}^{1}$ and $\mathcal{O}_{Qt}^{8}$ contribute to the matrix element in~\cref{eq:MgghOS4t}. This is a consequence of the colour structure of the diagrams and the on-shell kinematic configurations of the external states.

The matrix element for the Higgs-photon coupling ($\gamma\gamma H$) can be obtained by performing the replacements indicated in~\cref{app:matrixelements} on~\cref{eq:MgghOS4t} yielding
\begin{equation}
\label{eq:MgammagammahOS4t}
    \mathcal{M}^{\gamma \gamma H,\text{4t}}_{\text{OS}} = \left( c_{Qt}^{1} + c_F c_{Qt}^{8}\right)
    \mathcal{K}_{tG} \frac{Q_t e}{g_s}  \frac{1}{\Lambda^2}  \mathcal{M}_{t \gamma}^{\gamma \gamma H} + \left( c_{Qt}^{1} + c_F c_{Qt}^{8}\right) \frac{1}{\Lambda^2} \left( B_{ggH}+\mathcal{K}_{t \phi} \right) 
 \mathcal{M}_{\text{SM}}^{\gamma \gamma H}.
\end{equation}
We note that different combination of WCs enter in~\cref{eq:MgghOS4t} compared to~\cref{eq:MgammagammahOS4t}–first term of the former. The phenomenological consequences of this observation are discussed in the following section.

\section{Interpretation of SMEFT constraints in different $\gamma_5$ schemes}
\label{sec:interpretation}
In this section, we perform a simplified (toy) fit to assess the impact of the $\gamma_{5}$ prescription on the bounds on WCs from single–Higgs production. This fit uses the Higgs signal strength and its associated data from Ref.~\cite{CMS:2022dwd}.
We restrict our study to the dominant gluon–fusion channel and consider only total production rates, since measurements of the Higgs transverse-momentum spectrum are not yet available.
We use the same numerical inputs as in~\cref{sec:computationalsetup} and we set $\mu_R=\mu_F=m_H/2$ for Higgs production and $\mu_R =m_H$ for the partial widths~\cite{LHCHiggsCrossSectionWorkingGroup:2016ypw}.

In~\cref{tab:singleH} of~\cref{app:numerical_data} we list the numerical results for the single-Higgs production cross section, $\sigma$, and the Higgs partial width, $\Gamma$, computed in both $\gamma_{5}$ schemes using the formulae derived in the previous section.
We omit the $\mathcal{O}(\Lambda^{-4})$ terms, as they enter at one loop order higher than the SM–$\mathcal{O}(\Lambda^{-2})$ interference.

The theoretical signal strengths, $\mu^{\rm{Th}}$, used in the fit are defined as follows: 
\begin{equation}
\mu_{X}^{\rm Th}
\;=\;
\frac{\sigma_{\rm SMEFT}\,\mathrm{BR}(H\to X)_{\rm SMEFT}}
     {\sigma_{\rm SM}\,\mathrm{BR}(H\to X)_{\rm SM}},
\qquad
X \equiv [\gamma\gamma,\;W^{+}W^{-},\;ZZ,\;b\bar b,\;\tau^{+}\tau^{-},\;\mu^{+}\mu^{-}].
\label{eq:muthX}
\end{equation}
In~\cref{eq:muthX}, the same $K$-factor is used to account for higher-order corrections both in the SM and linear, $\order{\Lambda^{-2}}$, EFT contributions to the production cross section and so it cancels out in the signal strengths.

Concerning the branching ratios, $\mathrm{BR}(H\to X)$, in~\cref{eq:muthX}, four-top-quark operators modify only the loop–induced partial widths \(H\!\to\!gg\) and \(H\!\to\!\gamma\gamma\)–later denoted as $\Gamma_{gg}$ and $\Gamma_{\gamma\gamma}$, respectively. Consequently the total Higgs width, $\Gamma^{\rm{Tot}}$, changes and it multiplies all branching ratios by a common factor. Every theoretical signal strength in our fit therefore receives this
universal rescaling—except for \(\mu_{\gamma\gamma}^{\rm{Th}}\). Said channel is additionally affected by the process-specific four-top-quark contribution given in~\cref{eq:MgammagammahOS4t}. In particular, we have
\begin{equation}
\Gamma^{\text{Tot}} = \Gamma^{\text{Tot}}_{\rm{SM}} + K_{g g} \Gamma^{gg}_{\rm{int}} + K_{\gamma\gamma} \Gamma^{\gamma \gamma}_{\rm{int}},
\end{equation}
where $\Gamma^{gg}_{\rm{int}}$ and $\Gamma^{\gamma\gamma}_{\rm{int}}$ can be read from~\cref{tab:singleH}, whilst the $K-$factors $K_{gg}$ and $K_{\gamma \gamma}$ are obtained as the ratio between the SM best estimates~\cite{Inami:1982xt, Djouadi:1991tka, Spira:1995rr, Chetyrkin:1997iv, Baikov:2006ch, Actis:2008ts, Actis:2008ug,Zheng:1990qa,Djouadi:1990aj, Dawson:1992cy, Djouadi:1993ji, Melnikov:1993tj, Inoue:1994jq,Djouadi:1997yw} and the SM values in~\cref{tab:singleH}. The following are the values we employ~\cite{LHCHiggsCrossSectionWorkingGroup:2016ypw}: $\Gamma^{\text{Tot}}_{\rm{SM}} = 4.088 \, \text{MeV}$, $ K_{gg}=1.707$, $K_{\gamma\gamma}=0.913$.

We present here the theoretical signal strengths, expanded to linear order in the WCs, in the NDR scheme:  
\begin{equation} 
\label{eq:signalstrengthNDR}
    \begin{aligned}
    \mu_{\gamma \gamma}^{\text{Th}} &= 1 - 0.0159  \times c_{Qt}^{1} - 0.00239 \times c_{Qt}^{8} \\ 
    &-6.60 \times 10^{-5} (c_{Qt}^{1})^2
     -2.34 \times 10^{-5} (c_{Qt}^{8})^2
      -9.34 \times 10^{-5} c_{Qt}^{1}c_{Qt}^{8}
    , \\ 
    \mu_{\text{Y}}^{\text{Th}}&= 1 - 0.0186 \times c_{Qt}^{1} - 0.00606 \times c_{Qt}^{8}  \\
    &-1.47 \times 10^{-5} (c_{Qt}^{1})^2
    -1.17 \times 10^{-6} (c_{Qt}^{8})^2
    -8.39\times 10^{-6} c_{Qt}^{1}c_{Qt}^{8}
    , 
    \end{aligned}
\end{equation}
and the BMHV one: 
\begin{equation} \label{eq:signalstrengthBMHV}
    \begin{aligned}
    \mu_{\gamma \gamma}^{\text{Th}} &= 1 - 0.00451  \times c_{Qt}^{1} - 0.00601 \times c_{Qt}^{8}\\
    &-2.53 \times 10^{-6} (c_{Qt}^{1})^2
     -4.50\times 10^{-6} (c_{Qt}^{8})^2
      -6.75 \times 10^{-6} c_{Qt}^{1}c_{Qt}^{8}
    , \\ 
    \mu_{\text{Y}}^{\text{Th}}&= 1 - 0.00491 \times c_{Qt}^{1} - 0.00655 \times c_{Qt}^{8}\\
    &-5.66 \times 10^{-7} (c_{Qt}^{1})^2
     -1.01 \times 10^{-6} (c_{Qt}^{8})^2
      -1.51\times 10^{-6} c_{Qt}^{1}c_{Qt}^{8},
    \end{aligned}
\end{equation}
where $\mathrm{Y} \equiv [\;W^{+}W^{-},\;ZZ,\;b\bar b,\;\tau^{+}\tau^{-},\;\mu^{+}\mu^{-}]$.

\begin{figure}[ht]
    \centering
    \begin{subfigure}[t]{0.49\textwidth}
        \centering
        \includegraphics[width=\linewidth]{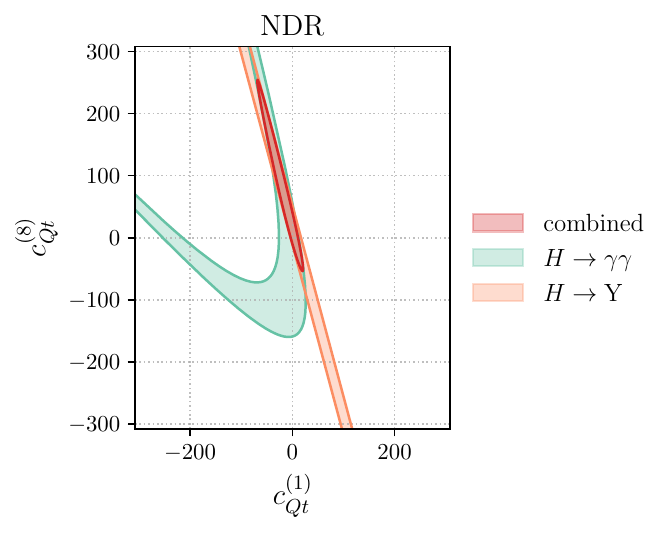}
    \end{subfigure}
    \hfill
    \begin{subfigure}[t]{0.49\textwidth}
        \centering
        \includegraphics[width=\linewidth]{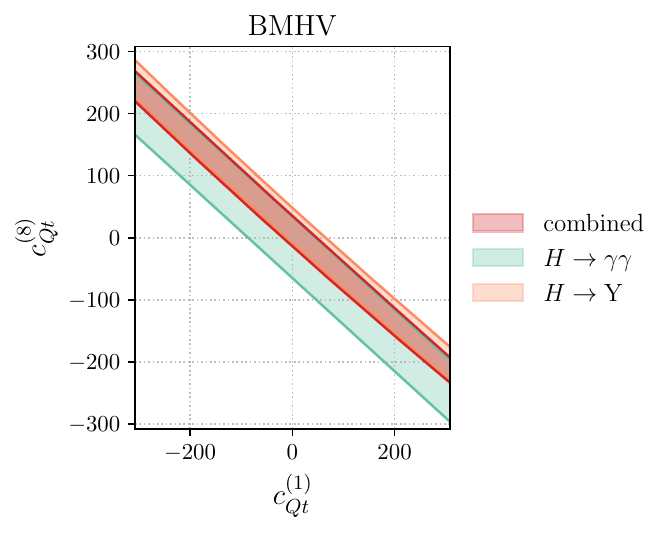}
    \end{subfigure}
    \caption{\(\Delta\chi^{2}\) contours for the single-Higgs production fit in the $(c_{Qt}^{1},\,c_{Qt}^{8})$ plane, shown for each decay channel, i.e. $\rm{Y}$ and $\gamma\gamma$ and for their combination, labelled as `combined'. The left (right) panel corresponds to the NDR (BMHV) $\gamma_{5}$ scheme. A pronounced flat direction emerges in the BMHV fit, whereas no such degeneracy appears in the NDR scheme.}
    \label{fig:BMHVvsNDR} 
\end{figure}

In~\cref{fig:BMHVvsNDR} we display the two-dimensional \(\Delta\chi^{2}\) contours for the single-Higgs production fit in the $(c_{Qt}^{1},\,c_{Qt}^{8})$ plane. This result highlights the scheme dependence introduced by the choice of $\gamma_{5}$-continuation prescription. The analysis is restricted to four-top-quark operators and the fit retains all linear, quadratic, and mixed (cross) terms in~\cref{eq:signalstrengthNDR,eq:signalstrengthBMHV}.  

As can be inferred from~\cref{fig:BMHVvsNDR}, the fit can distinguish between $c^{1}_{Qt}$ and  $c^{8}_{Qt}$ in NDR but finds a flat direction for $c^{1}_{Qt}+ c_F c^{8}_{Qt}$ in BHMV. This can be understood by inspecting~\cref{eq:MgghOS4t}: in the BMHV scheme, contributions of the type shown in~\cref{fig:4t_2loop_gtt} 
vanish, leaving a degeneracy in the WCs. Conversely, in the NDR scheme, they are non-vanishing and are proportional to the linear combination $c^{1}_{Qt} + \left(c_A -\dfrac{c_F}{2} \right) c^{8}_{Qt}$, which lifts the degeneracy.

As detailed in~\cref{app:matrixelements}, the WCs of operators entering at one-loop order depend on the scheme in such a way that they compensate for the scheme-dependent four-top-quark contribution.
This observation highlights that a $\gamma_{5}$-prescription–independent result can be obtained only when a sufficiently complete set of SMEFT operators is included, as demonstrated in Ref.~\cite{DiNoi:2023ygk}.  
If those additional operators are omitted, our fit—restricted to four-quark operators—must be interpreted differently in the NDR and BMHV schemes. 
Indeed, since the four-top operators behave differently in the two schemes, the corresponding fits probe distinct UV scenarios: either those that generate only four-top-quark operators in the NDR scheme, or those that generate only four-top-quark operators in the BMHV scheme. These represent two separate classes of UV models.

\section{Constraining four-quark operators: fit method, inputs and results}
\label{sec:fit}
\subsection{Fit method}
We analyse the impact of SMEFT operators on measured inclusive and differential cross sections, $\sigma_{\rm{Ex}}$, by performing individual and marginalised \(\chi^2\) fits. For an operator coefficient \(c_i\), the theoretical cross section, $\sigma_{\rm{Th}}$, in each bin, can be written as shown in~\cref{eq:theory_prediction}

When statistical and systematic uncertainties are provided separately by the experimental collaborations, the total experimental uncertainty, $\Delta_{\rm{Ex}}$, in each bin is determined by combining both uncertainties in a quadrature, i.e. assuming no correlation–the total experimental uncertainty provided directly by the collaborations is used when available. The normalisation, $\Delta_{\rm{Tot}}$, entering the test statistic is the quadrature sum of experimental and theoretical uncertainties, the latter being the QCD scale uncertainties of SM predictions. Conservatively, we choose that as the maximum of the scale uncertainty envelope. 

For each bin, the $\chi^2$ contribution is therefore calculated as
\begin{equation}
    \chi^2_{\text{bin}} = \left(\frac{\sigma_{\mathrm{Ex}} - \sigma_{\mathrm{Th}}}{\Delta_{\mathrm{Tot}}}\right)^2,
    \label{eq:chi2_bin}
\end{equation}
and the total $\chi^2$ is obtained by summing over all bins, as well as over all considered observables and processes: 
\begin{equation}
    \chi^2 = \sum_{\text{proc.}}\sum_{\text{obs.}}\sum_{\text{bins}} \chi^2_{\text{bin}}.
    \label{eq:chi2_total}
\end{equation}

Observables from different experiments are assumed to be uncorrelated. Finally, for the EWPO fit, we use the correlations between the different observables as quoted in Ref.~\cite{ALEPH:2005ab}.

\subsection{Fit inputs}
We discuss here the inputs for the processes included in our fit, which are summarised in~\cref{tab:process_overview}.
\begin{table}[ht]
    \centering
    \renewcommand{\arraystretch}{1.2}
    \resizebox{\textwidth}{!}{%
   \begin{tabular}{c c c c c c c c}
    \hline
    {Tag} & {$\sqrt{s},\,\mathcal{L}$} & {Final state} & {Observable} & {$n_{\mathrm{dat.}}$} & {Ref.(Ex)} & {Location/HEPData} & {Ref.(Th)} \\
    \hline
    \multicolumn{8}{c}{\boldmath{$pp \to t\bar{t}$}}\\
    \hline
    CMS$_{{t\bar{t}}}$ &  13 TeV, 137 fb$^{-1}$ & lepton+jets & $d\sigma/dm_{t\bar{t}}$ & 15 & ~\cite{CMS:2021vhb} & /Tab. 37 & ~\cite{CMS:2021vhb} \\
    ATLAS$_{{t\bar{t}}}$ & 13 TeV, 36 fb$^{-1}$ & lepton+jets & $d\sigma/dm_{t\bar{t}}$ & 9 & ~\cite{ATLAS:2019hxz} & /Tab. 617 & ~\cite{ATLAS:2019hxz}\\
    \hline
    \multicolumn{8}{c}{\boldmath{$pp \to t\bar{t}H$}}\\
    \hline
    ATLAS$_{{t\bar{t}H}}$ &  13 TeV, 140 fb$^{-1}$ & $H\to b\bar{b}$ & $d\sigma/{dp_T^H}$ & 6 & ~\cite{ATLAS:2024gth} & Fig. 3 in the ref./ & ~\cite{ATLAS:2022vkf}\\
    \hline
    \multicolumn{8}{c}{\boldmath{$pp \to t\bar{t}t\bar{t}$}}\\
    \hline
    CMS$_{4t}$ &  13 TeV, 138 fb$^{-1}$ & multi-leptons & $\sigma_{t\bar{t}t\bar{t}}^{\mathrm{tot.}}$ & 1 & ~\cite{CMS:2023ftu} & /Fig. 8 & ~\cite{vanBeekveld:2022hty}\\
    ATLAS$_{4t}$ &  13 TeV, 140 fb$^{-1}$ & multi-leptons & $\sigma_{t\bar{t}t\bar{t}}^{\mathrm{tot.}}$ & 1 & ~\cite{ATLAS:2023ajo} & /Tab. 17 & ~\cite{vanBeekveld:2022hty}\\
    \hline
    \multicolumn{8}{c}{\boldmath{$gg \to H$}}\\
    \hline
    CMS$_{ggH}$ & 13 TeV, 138 fb$^{-1}$ & $W^{+}W^-\,, ZZ\,, b\bar{b}\,, \tau^+\tau^-\,, \mu^{+}\mu^-$ & $\mu^{\rm{Ex}}_{\rm{Y}}$ & 1 & ~\cite{CMS:2022dwd} & /Tab. 12 & ~\cref{eq:signalstrengthNDR,eq:signalstrengthBMHV}\\
    CMS$_{ggH}$ & 13 TeV, 138 fb$^{-1}$ & $\gamma\gamma$ & $\mu^{\rm{Ex}}_{\gamma\gamma}$ & 1 & ~\cite{CMS:2022dwd} & /Tab. 12 & ~\cref{eq:signalstrengthNDR,eq:signalstrengthBMHV}\\
    \hline
    \multicolumn{8}{c}{\boldmath{$pp \to t\bar{t}b\bar{b}$}}\\
    \hline
    ATLAS$_{t\bar{t}b\bar{b}}$ & 13 TeV, 36.1 fb$^{-1}$ & lepton+jets & $\sigma_{t\bar{t}b\bar{b}}^{\mathrm{tot.}}$ & 1 & ~\cite{ATLAS:2018fwl} &/Tab. 5 & as in~\cite{Celada:2024mcf}\\
     CMS$_{t\bar{t}b\bar{b}}^{1}$ & 13 TeV, 35.9 fb$^{-1}$ & all-jets & $\sigma_{t\bar{t}b\bar{b}}^{\mathrm{tot.}}$ & 1 & ~\cite{CMS:2019eih} & Fig. 3 in the ref./ & as in~\cite{Celada:2024mcf}\\
     CMS$_{t\bar{t}b\bar{b}}^{2}$ & 13 TeV, 35.9 fb$^{-1}$ & dilepton & $\sigma_{t\bar{t}b\bar{b}}^{\mathrm{tot.}}$ & 1 & ~\cite{CMS:2020grm} & Tab. 4 in ref. (FPS)/ & as in~\cite{Celada:2024mcf}\\
     CMS$_{t\bar{t}b\bar{b}}^{3}$ & 13 TeV, 35.9, fb$^{-1}$ & lepton+jets & $\sigma_{t\bar{t}b\bar{b}}^{\mathrm{tot.}}$ & 1 & ~\cite{CMS:2020grm} & Tab. 4 in ref. (FPS)/ & as in~\cite{Celada:2024mcf}\\
    \hline
    \multicolumn{8}{c}{\bf{EWPO}}\\
    \hline
    EWPO$^{1}$ & $Z-$pole, / & $Z$ decays & $\Gamma_{Z}^{b\bar{b}}\,,R_{c}\,,R_{l}\,,R_{b}$ & 1 & ~\cite{ALEPH:2005ab} & Tabs. 7.1 and 8.4 in ref. & ~\cite{Brivio:2017bnu}\\
    EWPO$^{2}$ & $Z-$pole, / & $Z$ decays & $A_{b}\,,A_{b,\rm{FB}}$ & 1 & ~\cite{ALEPH:2005ab} & Tab. 8.4 in ref. & ~\cite{ALEPH:2005ab}\\
    \hline
    \end{tabular}}
   \caption{Summary of the inputs used in the fit. For each dataset we list, from left to right: (i) the dataset label; (ii) the centre-of-mass energy and integrated luminosity; (iii) the measured final state; (iv) the observable under study; (v) the number of data points; (vi) the experimental publication; (vii) the location of the experimental data (either in the publication or via its HEPData entry\protect\footnotemark); and (viii) the reference used for theoretical predictions.}
    \label{tab:process_overview}
\end{table}
\footnotetext{\url{https://www.hepdata.net/}}

\paragraph{\boldmath{$pp\to t\bar{t}$}} 
The SM predictions in the bins of ATLAS$_{t\bar{t}}$ are taken from the corresponding publication~\cite{ATLAS:2019hxz}, where MC simulations were generated using \texttt{Powheg-Box} v2~\cite{Frixione:2007nw,Nason:2004rx,Frixione:2007vw,Alioli:2010xd} interfaced with \texttt{Pythia} 8.210~\cite{Sjostrand:2007gs}. The cross-section normalisation of these MC samples is set to the NNLO + next-to-next-to-leading-logarithmic (NNLL) QCD prediction~\cite{Czakon:2011xx,Beneke:2011mq,Cacciari:2011hy,Barnreuther:2012wtj,Czakon:2012zr,Czakon:2012pz,Czakon:2013goa}, as indicated explicitly in Table~1 of the publication. Predictions and measurements from ATLAS$_{t\bar{t}}$ are normalised to the total cross section of $832^{+20}_{-29}\,\mathrm{pb}$, as reported therein.

The SM predictions in the bins of CMS$_{t\bar{t}}$ are also taken from the corresponding publication~\cite{CMS:2021vhb}. The cross-section normalisation of these MC samples is set to the NNLO in the strong coupling constant including the resummation of NNLL soft-gluon terms calculated with TOP++ (version 2.0)~\cite{Czakon:2011xx}.
This amounts to a normalisation factor, i.e. the inclusive $t\bar{t}$ production cross section, of $832^{+40}_{-46}\,\mathrm{pb}$. The SM theoretical uncertainties are taken from the respective publications.

\paragraph{\boldmath{$pp\to t\bar{t}H$}}
We use the SM predictions and the associated uncertainties provided in the analysis of Ref.~\cite{ATLAS:2022vkf}. We use the experimental data from the most recent measurement of Ref.~\cite{ATLAS:2024gth} reporting a measured total cross section of $411^{+24\%}_{-22\%}\,\mathrm{fb}$. 
The SM differential predictions~\cite{LHCHiggsCrossSectionWorkingGroup:2016ypw,Lindert:2018iug,Djouadi:2018xqq,Bonetti:2018ukf,Dulat:2018rbf,Harlander:2018yio,Cacciari:2015jma} extracted from Ref.~\cite{ATLAS:2022vkf} are in good agreement with our own results presented in~\cref{tab:tth_pth_pred} of~\cref{app:numerical_data}.

\paragraph{\boldmath{$pp\to t\bar{t}t\bar{t}$}}
For both datasets, we adopt the SM cross section of Ref.~\cite{vanBeekveld:2022hty}, computed at NLO (QCD+EW)+NLL accuracy, yielding $13.37^{+7.77\%}_{-13.3\%}\,\mathrm{fb}$.

\paragraph{\boldmath{$pp\to t\bar{t}b\bar{b}$}} We extract the experimental measurements and theoretical predictions—for \emph{both} the SM–assigned a 10\% theoretical uncertainty– and the SMEFT—from Ref.~\cite{Celada:2024mcf}.\footnote{The SMEFiT database can be found here: \url{https://github.com/LHCfitNikhef/smefit_database/}}

\paragraph{\boldmath{$gg\to H$}} The analysis presented in Ref.~\cite{CMS:2022dwd} reports signal-strength modifiers, \(\mu_{i}^{\rm Ex}\), categorised through their decay modes $i$, with uncertainties representing the total experimental error, combining both systematic and statistical contributions. These results are summarised as follows:
\begin{equation}
    \begin{aligned}
    \mu^{\rm{Ex}}_{\mu\mu} &= 0.33^{+0.74}_{-0.70}\,, &
    \mu^{\rm{Ex}}_{\tau\tau} &= 0.66^{+0.21}_{-0.21}\,, &
    \mu^{\rm{Ex}}_{ZZ} &= 0.93^{+0.14}_{-0.13}\,, \\[0.2cm]
    \mu^{\rm{Ex}}_{WW} &= 0.90^{+0.11}_{-0.10}\,, &
    \mu^{\rm{Ex}}_{bb} &= 5.31^{+2.97}_{-2.54}\,, &
    \mu^{\rm{Ex}}_{\gamma\gamma} &= 1.08^{+0.12}_{-0.11}\,.
    \end{aligned}
    \label{eq:mu_ggh}
\end{equation}
The results reported in~\cref{eq:mu_ggh} are of the gluon-fusion production mode and constitute the experimental input for our fit. Given that  gluon fusion is directly sensitive to the operators we consider and is the dominant production mode for single-Higgs production, we do not expect significant sensitivity from other production modes.We set the SM prediction to unity.

\paragraph{\textbf{EWPO}} Our predictions are obtained at the two‐loop level leveraging the work of Ref.~\cite{Haisch:2024wnw} as discussed in~\cref{sec:EWPO} and in~\cref{app:ewpo}. Experimental measurements and correlations are taken from Ref.~\cite{ALEPH:2005ab}. SM predictions for all observables apart from $A_{b}$ and $A_{b,\rm{FB}}$ are extracted from Table 2 of Ref.~\cite{Brivio:2017bnu}. SM predictions of $A_{b}$ and $A_{b,\rm{FB}}$ are taken directly from Table 8.4 in Ref.~\cite{ALEPH:2005ab}.

\subsection{Fit results} 
\label{sec:results}
\begin{figure}[ht]
    \centering
    \includegraphics[scale=0.55]{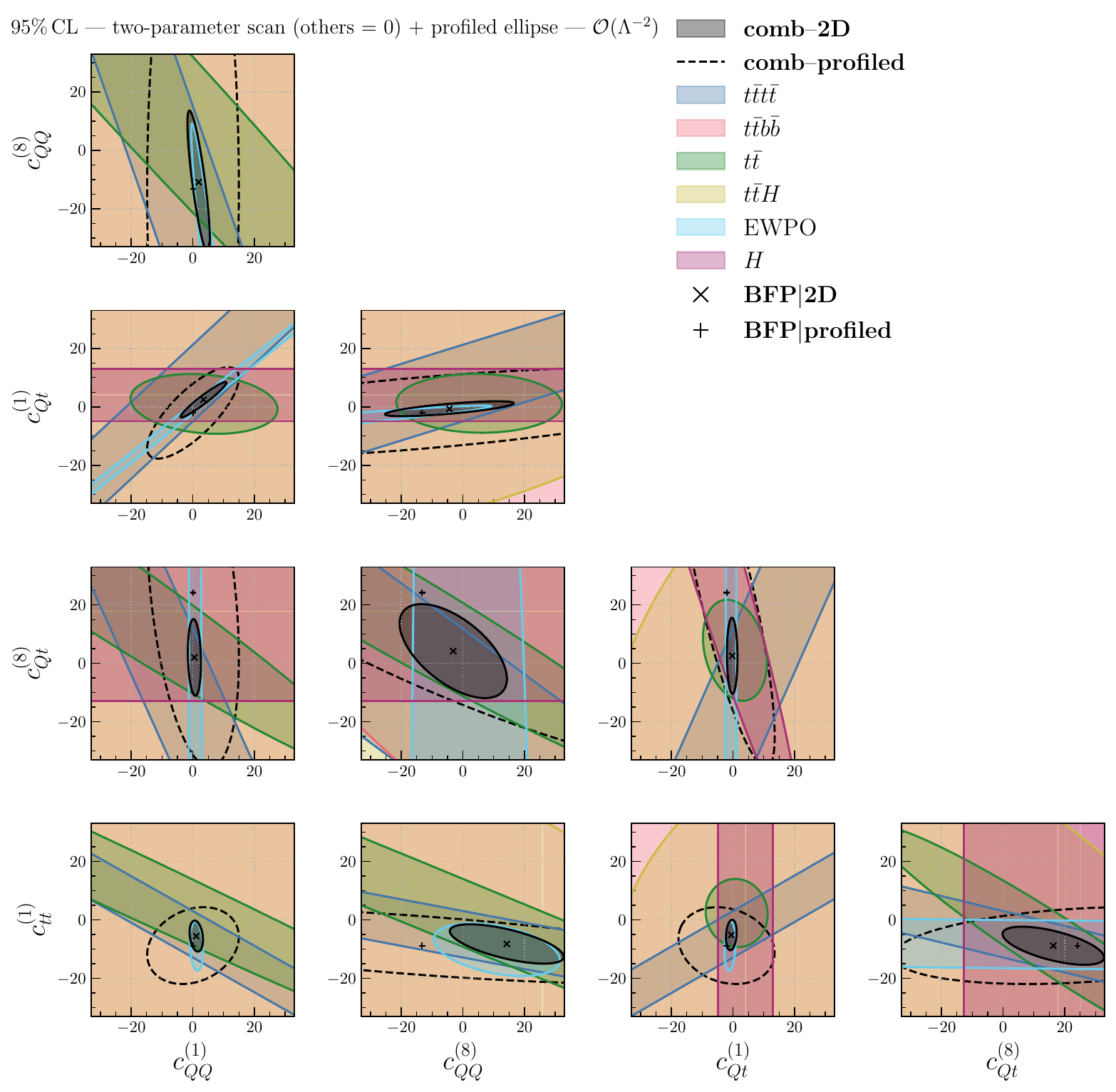}
    \caption{Two-dimensional fits for the four–heavy operator coefficients. Shown are the constraints from each set of observables separately and their combination. Only linear terms, \(\mathcal{O}(\Lambda^{-2})\), in the EFT parametrisation are included. The best-fit point (BFP) for the combined fit is indicated for both the two-parameter scan and the profiled scan.}
    \label{fig:cont_int}
\end{figure}

\begin{figure}[ht]
    \centering
    \includegraphics[scale=0.55]{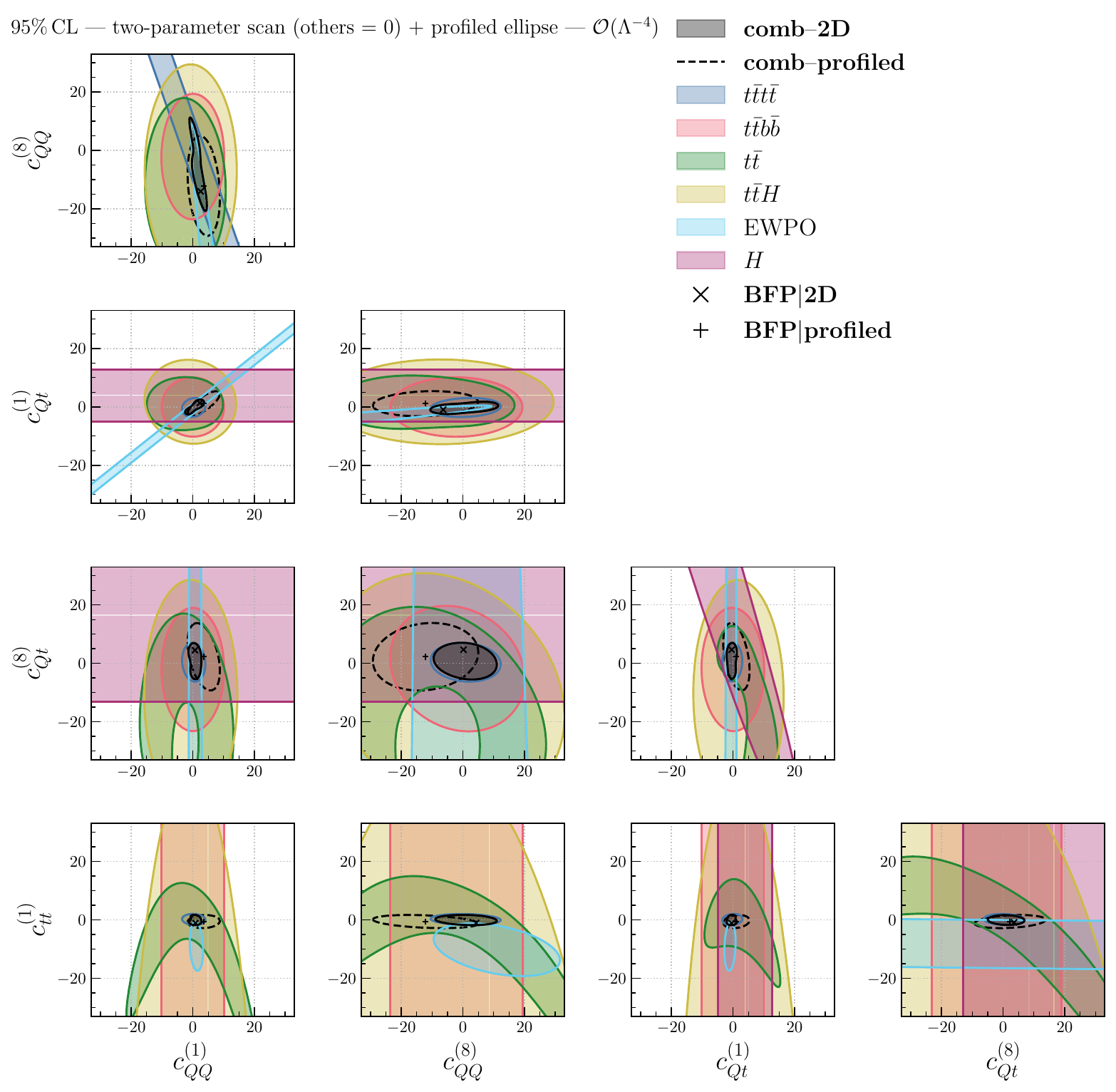}
    \caption{Same as~\cref{fig:cont_int}, but including quadratic terms, \(\mathcal{O}(\Lambda^{-4})\), in the EFT parametrisation.}
    \label{fig:cont_sq}
\end{figure}

\Cref{fig:cont_int} shows the two-dimensional \(\Delta\chi^{2}\) contours at 95\%~CL for the case where only linear EFT terms are included whilst~\cref{fig:cont_sq} corresponds to the scenario where quadratic contributions are also taken into account. 
In both panels, we display the contours for the combined fit (all processes) under two distinct scenarios: one in which only the two WCs of interest are varied (black solid-line contour, labelled as `comb-2D' in the plots) with all other coefficients fixed to zero, and one in which those two coefficients are scanned while the remaining WCs are profiled (black dashed-line contour, labelled as `comb-profiled' in the plots). The best-fit point (BFP) is indicated in each case. In~\cref{tab:all-bounds} of~\cref{app:numerical_data} we list all the  95\%~CL bounds obtained on each of the five WCs. Finally, we note that all our upcoming results–including $gg \to H$ results– are obtained using the NDR scheme.

We observe that, in the linear case of~\cref{fig:cont_int}, the comb-2D fit yields smaller and differently shaped contours than those obtained from the individual channels. This illustrates that the inclusion of additional production modes provides complementary information and strengthens the overall constraints. By contrast, the comb-profiled contours display significantly weaker constraining power in comparison to the comb-2D ones, underscoring the impact of the profiled coefficients.
Examining at individual datasets, we observe that most are plagued by flat directions; four-top-quark production, for example, shows flat directions for all coefficient pairs we consider. Top-quark pair production in association with a Higgs and $t\bar{t}b\bar{b}$ seem to exhibit the least constraining power among the different processes. Inclusive Higgs production probes only a subset of the coefficients and also has no significant impact on the final combination. 

Furthermore, we observe a sizeable impact of the two-loop corrections in our EWPO fit—particularly when compared with the corresponding one-loop contours of Ref.~\cite{Degrande:2024mbg}. These effects are driven mainly by the non-negligible two-loop contributions to the asymmetry observables \(A_b\) and \(A_{b,\mathrm{FB}}\); see~\cref{app:ewpo} for a detailed comparison between the one- and two-loop results for these quantities.
We find that the combination of EWPO, top-quark pair and four-top-quark production significantly reduces the allowed parameter space. For some pairs we note that the contours extracted from the EWPO do not include the SM at the 95$\%$ CL. This is related to the well-known discrepancy between the measurement of \(A_{b,\mathrm{FB}}\) and its SM prediction \cite{ALEPH:2005ab}. 

At the quadratic level of~\cref{fig:cont_sq}, the combined fit yields markedly tighter bounds, driven predominantly by the four–top-quark measurements. The comb-2D contour lies very close to the four–top-quark one, with modest additional tightening from the EWPO. Moreover, the discrepancy between the comb-profiled and comb-2D contours is less pronounced than in the linear case. The most significant reduction in the allowed parameter space appears in the $(c^{(8)}_{QQ}$–$c^{(1)}_{QQ})$ plane, where the four–top-quark channel alone exhibits an almost blind direction; here the EWPO combination is essential to lift the degeneracy.

\section{Conclusion}
\label{sec:conclusions}
In this work we explored how various classes of observables can shed light on the dimension-six four-heavy-quark operators. These operators are notoriously difficult to constrain: global EFT fits typically leave them essentially undetermined at the linear level. To overcome this, several studies have examined loop-induced effects in specific observables, which offer complementary sensitivity to these otherwise elusive interactions.

In particular, we have explored the tree‐level contributions of the four‐heavy operators to four‐top-quark production at the LHC; their one‐loop contributions to top‐quark pair production and to associated Higgs production–both inclusive and differentially; the two‐loop contributions to single-Higgs production in gluon fusion and its subsequent decays; and the one‐ and two‐loop contributions to EWPO. We presented results for all these channels at linear and, where available, quadratic terms in the EFT expansion. 

In the case of single-Higgs production via gluon fusion and its subsequent decays, we scrutinised the dependence of the results on the choice of the prescription for the continuation of $\gamma_{5}$ to $d =4-2\epsilon$. 
We show that this scheme dependence numerically propagates into the bounds on the WCs. Therefore, one must exercise caution when interpreting these limits.
Indeed, by restricting the analysis to four-top-quark operators, this assumption leads to a different interpretation of the fit in the BMHV scheme compared with NDR. Establishing a coherent correspondence between the two schemes therefore necessitates the inclusion of additional operators. We defer a full global fit in the BMHV scheme to future work, as our restricted fit clearly calls for a global picture including a full basis of operators.

Finally, we performed a fit combining experimental data from the LHC and LEP, incorporating all relevant theoretical and experimental uncertainties. This allowed us to identify the most sensitive observables and to elucidate the complementarity between tree- and loop-level probes. In the linear fit, the synergy between heavy‐quark production and EWPO is key to reducing the allowed parameter space. In the quadratic fit, the strongest constraints arise from four-top-quark production, with the EWPO further reducing the allowed parameter space and lifting degeneracies where present. Moreover, we observed non-negligible two-loop effects in the EWPO fit, predominantly driven by the sizeable two-loop corrections to the asymmetry observables. 

In conclusion, we note that a comprehensive assessment of the impact of the observables considered in this work requires their inclusion in a global fit together with the complete set of relevant operators. It is also worthwhile to investigate UV-complete theories which, upon matching onto the EFT, reproduce the operator basis examined here. 

\section*{Acknowledgements}
HF thanks Uli~Haisch for helpful discussions and for clarifying technical details of Ref.~\cite{Haisch:2024wnw}; Víctor~Miralles for discussions on the fit method; Marion~Thomas for her input on the EWPO implementation; and Giuseppe~Ventura for discussions on the $\gamma_5$ prescriptions. SDN and RG would like to thank Pablo Olgoso for various discussions on $\gamma_5$. EV thanks Gauthier~Durieux for valuable discussions on the SMEFT‐basis arguments presented in~\cref{sec:computationalsetup}. The work of HF and EV is supported by the European Research Council (ERC) under the European Union’s Horizon 2020 research and innovation programme (Grant Agreement No. 949451) and by a Royal Society University Research Fellowship (Grant URF/R1/201553). The work of RG is supported in part by the Italian MUR Departments of Excellence grant 2023-2027 ``Quantum Frontiers”, by the INFN Iniziativa Specifica APINE and by the University of Padua under the 2023 STARS Grants@Unipd programme (Acronym and title of the project: HiggsPairs – Precise Theoretical Predictions for Higgs pair production at the LHC).
The research of SDN and MV is supported by the Deutsche Forschungsgemeinschaft (DFG, German Research Foundation) under grant 396021762 — TRR 257 “Particle Physics Phenomenology
after the Higgs Discovery”.

\appendix
\section{Matrix elements for \boldmath{$ggH$} and \boldmath{$\gamma\gamma H$}}  
\label{app:matrixelements}
In this appendix we provide the explicit expression for the auxiliary quantities appearing in~\cref{eq:MgghOS4t,eq:MgammagammahOS4t}. We also include the contribution from the operators
\begin{equation}
\begin{aligned}
    \mathcal{O}_{t \phi} \equiv \mathcal{Q}_{u \phi}^{(33)} = (\bar{q}_3 \tilde{\phi} u_3 ) \left( \phi^\dagger \phi \right),&\quad \mathcal{O}_{tG} \equiv \mathcal{Q}_{uG}^{(33)} = \bar{q}_3 \tilde{\phi} \sigma^{\mu\nu} T^A u_3 G^A_{\mu\nu}, \\ 
    \mathcal{O}_{\varphi G} \equiv \mathcal{Q}_{\varphi G} =& G^{A,\mu \nu}G^A_{\mu \nu}\left( \phi^\dagger \phi \right),
\end{aligned}
\end{equation}
which are required to obtain a result that is independent of the $\gamma_5$ continuation scheme. We note that the operator $\mathcal{O}_{\varphi G}$ is introduced to renormalise the one-loop contribution from $\mathcal{O}_{tG}$. We note that the results presented here employ the $\overline{\text{MS}}$ renormalisation scheme for the WCs and the on-shell scheme for the top-quark mass, in contrast to Ref.~\cite{DiNoi:2023ygk}, where all parameters were renormalised in the MS scheme.

The matrix element for the process \(gg \to H\) (or, equivalently, \(H \to gg\)) reads 
\begin{equation}
\label{eq:MgghOS}
    \begin{aligned} 
    \mathcal{M}^{ggH}_{\text{OS}} =& \frac{c_{\phi G }}{\Lambda^2} \mathcal{M}_{\phi G}^{ggH} + \left[   c_{tG} + 
    \left( c_{Qt}^{1} + \left(c_F-\dfrac{c_A}{2} \right) c_{Qt}^{8}\right)
    \mathcal{K}_{tG}   \right]  \frac{1}{\Lambda^2} \mathcal{M}_{tG}^{ggH} \\ +& \left[ 1+ \left( c_{Qt}^{1} + c_F c_{Qt}^{8}\right) \frac{1}{\Lambda^2} \left( B_{ggH}+ \mathcal{K}_{t \phi} \right) 
    -\frac{v^3}{\sqrt{2} m_t \, \Lambda^2} c_{t\phi}{}
    \right]\mathcal{M}_{\text{SM}}^{ggH},
    \end{aligned}
\end{equation}
where \(c_A\) and \(c_F\) are the \(\mathrm{SU}(N)\) Casimir invariants in the adjoint and fundamental representations, respectively. For \(\mathrm{SU}(3)_\mathrm{C}\) they take the values \(c_A = 3\) and \(c_F = 4/3\). The factor \(B_{ggH}\) is defined in~\cref{eq:F4t}.

The matrix elements entering~\cref{eq:MgghOS} are  
\begin{align}
    \mathcal{M}_{tG}^{ggH} &= - T_F\frac{g_s m_t }{\sqrt{2} \pi^2} L^{\mu_1 \mu_2}  \epsilon_{\mu_1}(p_1) \epsilon_{\mu_2} (p_2) \delta^{A_1A_2}   \times \\ 
        & \;   \left(\frac{m_t^2}{m_H^2} \log^2 \left(\frac{\beta-1}{\beta + 1} \right) + \sqrt{1- \frac{4 m_t^2}{m_H^2}} \log\left(\frac{\beta-1}{\beta + 1} \right)  
        + 2 \log\left(\frac{{\mu}_{R}^2}{m_t^2} \right) + 1 \right), \nonumber \\
        \mathcal{M}_{\text{SM}}^{ggH} & = T_F\frac{ g_s^2 m_t^2}{4 \pi^2 v m_H^2}L^{\mu_1 \mu_2}  \epsilon_{\mu_1}(p_1) \epsilon_{\mu_2} (p_2)  \delta ^{A_1A_2}\times \left(\frac{m_H^2-4m_t^2}{m_H^2}  \log^2\left(\frac{\beta -1}{\beta +1}\right)-4\right)
        , \\ 
    \mathcal{M}^{ggH}_{\phi G} &=  - 4 i v  L^{\mu_1 \mu_2}  \epsilon_{\mu_1}(p_1) \epsilon_{\mu_2} (p_2) \delta^{A_1A_2}.      
\end{align}
Here \(T_F\) is the Dynkin index of the fundamental representation of \(\mathrm{SU}(N)\).
The Lorentz structure of the amplitude is 
\begin{equation} 
\label{eq:Lorentz}
    L^{\mu_1 \mu_2} = (m_H^2/2   ~g^{\mu_1 \mu_2}- p_1^{\mu_2}p_2^{\mu_1} ),
\end{equation}
with \(p_1\) and \(p_2\) being the gluon momenta. The indices \(A_1\) and \(A_2\) denote the gluon colour indices.
 
The scheme dependence arising from the two-loop contributions of four-top-quark operators, parametrised by $\mathcal{K}_{tG}$, and $\mathcal{K}_{t \varphi}$ in~\cref{eq:Kggh}, can be compensated by assuming the WCs of operators entering at one loop order are scheme dependent. In particular, the relations 
\begin{align}
    c_{t \phi }^{\text{NDR}}  &= c_{t \phi }^{\text{BMHV}} +
    \left( c_{Qt}^{1} + c_F c_{Qt}^{8}\right)  \frac{y_{t}(\lambda -y_{t}^2)}{8\pi^2 },   \label{eq:shiftsctphi}\\
    c_{tG}^{\text{NDR}}  &=c_{tG}^{\text{BMHV}} - \left( c_{Qt}^{1} + \left(c_F-\dfrac{c_A}{2} \right)c_{Qt}^{8}\right)
    \frac{y_t g_s}{16 \pi^2} , \label{eq:shiftsctG}
\end{align}
render the prediction in~\cref{eq:MgghOS} scheme independent. In these expressions, $y_t$ is the top-quark Yukawa coupling and $\lambda = m_H^2/(2 v^2)$. We note that some of the shifts depend on the strong coupling constant, $g_{s}$, and thus on the renormalisation scale. This scale dependence must be accounted for when a dynamical scale choice is employed in the calculation.

We now present the matrix element for the Higgs-photon coupling. 
We use $F^{\mu \nu}$ to denote the photon field strength tensor and introduce the operators $\mathcal{O}_{t \gamma}{}$ $= (\bar{t}_L   \sigma^{\mu \nu} t_R) \frac{H+v}{\sqrt{2}} F_{\mu \nu}$ and $\mathcal{O}_{\phi \gamma}{} =Hv F^{\mu \nu }F_{\mu \nu }$. These operators are not part of the Warsaw basis, as they are defined directly in terms of the physical fields in the broken phase.
Their expression in Warsaw-basis operators can be found in Refs.~\cite{rosetta,Higgsbasis}. 
We finally obtain
\begin{equation}
\label{eq:MgammagammahOS}
    \begin{aligned} 
    \mathcal{M}^{\gamma \gamma H}_{\text{OS}} =& \mathcal{M}_{\text{SM},W}^{\gamma \gamma H}+ \frac{c_{\phi \gamma }{}}{\Lambda^2} \mathcal{M}_{\phi \gamma}+ \left[  c_{ t \gamma } + 
    \left( c_{Qt}^{1} + c_F c_{Qt}^{8}\right)
    \mathcal{K}_{tG} \frac{Q_t e}{g_s}   \right] \frac{1}{\Lambda^2}  \mathcal{M}_{t \gamma}^{\gamma \gamma H} \\ +& \left[ 1+ \left( c_{Qt}^{1} + c_F c_{Qt}^{8}\right) \frac{1}{\Lambda^2} \left( B_{ggH}+\mathcal{K}_{t \phi} \right) 
    -\frac{v^3}{\sqrt{2} m_t \, \Lambda^2} c_{t\phi}
    \right]\mathcal{M}_{\text{SM}}^{\gamma \gamma H}.
    \end{aligned}
\end{equation}

The one-loop matrix element for \(H \to \gamma\gamma\) can be obtained from that for \(H \to gg\) by making the substitutions $g_s \to e Q_t$ and $T_F \delta^{A_1 A_2} \to N_C $ in $\mathcal{M}_{tG}^{ggH},~\mathcal{M}_{\mathrm{SM}}^{ggH}$, where $e$ is the electric charge of the electron and $Q_t=2/3$ is the quantised charge of the top quark. We denote the matrix elements of the Higgs boson decay into photons as $\mathcal{M}_{t\gamma}^{\gamma \gamma H},~\mathcal{M}_{\mathrm{SM}}^{\gamma \gamma H}$. The tree-level insertion of $\mathcal{O}_{\varphi \gamma}{}$ is given by $\mathcal{M}^{\gamma \gamma H}_{\phi \gamma } =  - 4 i v  L^{\mu_1 \mu_2}  \epsilon_{\mu_1}(p_1) \epsilon_{\mu_2} (p_2)$. Regarding the two-loop matrix elements, each four-top-quark operator insertion generates two different colour contractions. We are able to obtain our result from the Higgs-gluon coupling since in the diagrams in~\cref{fig:4t_2loop}, the only non-vanishing term features a single Dirac trace, allowing the colour structures to be identified unambiguously. 

For completeness, the SM contribution to $H \to \gamma \gamma$ with $W$-boson loops reads 
\begin{equation}
    \begin{aligned}
    \mathcal{M}_{\text{SM},W}^{\gamma \gamma H} & =\frac{e^2 }{4 \pi ^2 v } {\left( \frac{6m_W^2}{m_H^2}+\left( \frac{6m_W^4}{m_H^4}-\frac{3m_W^2}{m_H^2}\right) \log
   ^2\left(\frac{\beta_W -1}{ \beta_W+1}\right)+1\right)},
    \end{aligned}
\end{equation}
where $\beta_W=\sqrt{1- \frac{4 m_W^2}{m_H^2}}$.

To render the Higgs decay into photons scheme–independent, we must employ~\cref{eq:shiftsctphi} with the relation analogous to~\cref{eq:shiftsctG}, namely
\begin{equation}
    c_{t\gamma}^{\text{NDR}}  =c_{t \gamma}^{\text{BMHV}} -e Q_t \frac{ y_t  \left(c^{1}_{Qt} + c_F c^{8}_{Qt} \right)}{8  \pi ^2 }. \label{eq:shiftsctgamma}
\end{equation}

\section{Numerical predictions}
\label{app:numerical_data}
\Cref{tab:tth_pth_pred} presents the differential \(p_T^H\) spectrum in the ATLAS \(t\bar tH\) bins, whilst~\cref{tab:all_pred_incl} lists the inclusive cross sections for all top-quark processes.
SMEFT results are split into interference, quadratic, and cross terms, with total-rate \(K\)-factors given. All WCs are set to unity and \(\Lambda = 1\;\text{TeV}\).
Predictions are quoted within their QCD scale uncertainties and MC statistical errors. Predictions which are compatible with zero within a MC error of 2$\sigma$ or greater are omitted and denoted by the `-'. The `x' indicates no contribution from the corresponding operator. 

\Cref{tab:singleH} lists our numerical predictions for single–Higgs production and for Higgs decays into gluons and photons. The quoted SM value is the LO result with only top- and $W$-boson loops included. Quadratic SMEFT terms are strongly suppressed relative to the linear ones and are therefore not shown.
\begin{table}[ht]
    \centering
    \scriptsize 
    \renewcommand{\arraystretch}{1.2} 
    \setlength{\tabcolsep}{3pt} 
    \input{table_pred_pTH}
   \caption{Differential \(p_T^H\) predictions in the SM and SMEFT in the $t\bar{t}H$ process. The cross sections are reported as the integral within the bin range. The uncertainties quoted correspond, respectively, to the Monte Carlo statistical uncertainty and the QCD scale uncertainty from a 9-point variation.}
    \label{tab:tth_pth_pred}
\end{table}
\begin{landscape}
\begin{table}[ht]
    \centering
    \scriptsize 
    \renewcommand{\arraystretch}{1.05} 
    \setlength{\tabcolsep}{3pt} 
    \input{table_pred_incl_all}
    \caption{Inclusive predictions for top-quark processes in the SM and SMEFT.}
    \label{tab:all_pred_incl}
\end{table}
\end{landscape}
\begin{table}[ht]
  \centering
  \scriptsize
  \renewcommand{\arraystretch}{1.15}
  \setlength{\tabcolsep}{4pt}
  \captionsetup[subtable]{font=small,labelfont=bf}
  \begin{subtable}[t]{0.32\linewidth}
    \centering
    \begin{tabular}{@{}lc@{}}
      \toprule
     
       & $\sigma_{\rm int}\,[\mathrm{pb}]$ \\ \midrule
      \multicolumn{2}{c}{$\mathcal{O}(c_i/\Lambda^{2})$}\\[2pt]
      $c_{Qt}^{1}$ (NDR)  & $-0.3203$ \\
      $c_{Qt}^{8}$ (NDR)  & $-0.1033$ \\
      $c_{Qt}^{1}$ (BMHV) & $-0.0830$ \\
      $c_{Qt}^{8}$ (BMHV) & $-0.1106$ \\
      \midrule 
       $\sigma_{\rm SM}$ & 16.51 pb \\
      \bottomrule
    \end{tabular}
    \caption{$gg\to H$ cross section}
    \label{tab:singleH_ggF}
  \end{subtable}\hfill
  \begin{subtable}[t]{0.32\linewidth}
    \centering
    \begin{tabular}{@{}lc@{}}
      \toprule
       & $\Gamma_{\rm int}\,[\mathrm{MeV}]$ \\ \midrule
      \multicolumn{2}{c}{$\mathcal{O}(c_i/\Lambda^{2})$}\\[2pt]
      $c_{Qt}^{1}$ (NDR)  & $-1.912\times10^{-3}$ \\
      $c_{Qt}^{8}$ (NDR)  & $-4.810\times10^{-4}$ \\
      $c_{Qt}^{1}$ (BMHV) & $-2.781\times10^{-4}$ \\
      $c_{Qt}^{8}$ (BMHV) & $-3.708\times10^{-4}$ \\ \midrule 
      $\Gamma_{\rm SM}$ & 0.1960 MeV \\
      \bottomrule
    \end{tabular}
    \caption{$H \to gg$ partial width}
    \label{tab:singleH_hgg}
  \end{subtable}\hfill
  \begin{subtable}[t]{0.32\linewidth}
    \centering
    \begin{tabular}{@{}lc@{}}
     \toprule
       & $\Gamma_{\rm int}\,[\mathrm{MeV}]$ \\ \midrule
      \multicolumn{2}{c}{$\mathcal{O}(c_i/\Lambda^{2})$}\\[2pt]
      $c_{Qt}^{1}$ (NDR)  & $ 2.798\times10^{-5}$ \\
      $c_{Qt}^{8}$ (NDR)  & $ 3.731\times10^{-5}$ \\
      $c_{Qt}^{1}$ (BMHV) & $ 4.069\times10^{-6}$ \\
      $c_{Qt}^{8}$ (BMHV) & $ 5.426\times10^{-6}$ \\ \midrule 
      $\Gamma_{\rm SM}$ & $1.016 \times 10^{-2}$ MeV \\
      \bottomrule
    \end{tabular}
    \caption{$H \to \gamma\gamma$ partial width}
    \label{tab:singleH_hgammagamma}
  \end{subtable}
  \caption{Linear EFT contributions and SM values for single-Higgs inclusive production cross section in gluon-fusion channel (\cref{tab:singleH_ggF}) and Higgs decays (\cref{tab:singleH_hgg,tab:singleH_hgammagamma}), parametrised as in~\cref{eq:theory_prediction}. WCs are set to unity with $\Lambda = 1\,$TeV. The SM values correspond to the LO one-loop result including only the dominant top- and $W$-loops. Results are presented in both the NDR and the BMHV schemes.}
  \label{tab:singleH}
\end{table}

\section{EWPO}
\label{app:ewpo}
We employ the relations in~\cref{eq:ST_relations_app} where the WCs \(C_{HD}\) and \(C_{HWB}\) are expressed in terms of the shifts to the oblique parameters \(S\) and \(T\)~\cite{Dawson:2019clf} and in terms of the shifts of the Weinberg angle, $s_\theta$, and the effective couplings~\cite{Brivio:2017bnu}. The two-loop contributions to the oblique parameters \(\Delta S\) and \(\Delta T\) have been computed in and are taken from Ref.~\cite{Haisch:2024wnw}.
\begin{equation}
    \begin{aligned}
    C_{HD} &= -\frac{2\,\alpha\,\Delta T}{v^2}, \quad
    C_{HWB} = \frac{\alpha\,\Delta S}{4\, c_{\theta}\, s_{\theta}\, v^2}, \\
    \delta s_\theta^{2} &= \frac{m_W^2\, C_{HD}}{2\sqrt{2}\, G_F\, m_Z^2} 
    + \frac{m_W\, C_{HWB}}{\sqrt{2}\, G_F\, m_Z\,}\sqrt{1 - \frac{m_W^2}{m_Z^2}}, \quad
    \delta g_Z = -\frac{C_{HD}}{4 \sqrt{2}\, G_F}.
\end{aligned}
\label{eq:ST_relations_app}
\end{equation}
Using~\cref{eq:ST_relations_app} and substituting into the shifts in vector and axial-vector couplings shown in~\cref{eq:mod_coupling_app}—which can be found in Ref.~\cite{Brivio:2017bnu}—we obtain the modified vector and axial–vector couplings due to the effective operators:
\begin{equation}
\begin{aligned}
    \delta g_V^f &= \delta g_Z\,g_V^f + Q^f\,\delta s_\theta^2\,,\\
    \delta g_A^f &= \delta g_Z\,g_A^{f}\,,
\end{aligned}
\label{eq:mod_coupling_app}
\end{equation}
where $g_{V}^{f}=T_{3}/2 - Q^{f}s_{\theta}^{2}$ and  $g_{A}^{f}=T_{3}/2$, where $T_{3}$ is weak isospin and $Q^{f}$ is the electric charge. We adopt the conventions of Ref.~\cite{Brivio:2017bnu} and use the following definitions of the EWPO:
\begin{equation}
\setlength{\jot}{4pt}
\begin{aligned}
\Gamma_i               &= \frac{\sqrt{2}\, G_F\, m_Z^{3}\,N_c}{3\pi}\,
                               \bigl(|g_V^{\,i}|^{2}+| g_A^{\,i}|^{2}\bigr),\\
\Gamma_{\mathrm{had}}  &= \sum_{q=u,d,c,s,b}\Gamma_q, \hspace{1cm}
 R_c                   = \frac{\Gamma_c}{\Gamma_{\mathrm{had}}},\quad
 R_b                   = \frac{\Gamma_b}{\Gamma_{\mathrm{had}}},\quad
 R_\ell                = \frac{\Gamma_{\mathrm{had}}}{\Gamma_\ell}.
\end{aligned}
\end{equation}
For the numerical analysis we adopt the following input parameters:
\begin{equation}
\label{eq:inputs}
\setlength{\jot}{4pt}
\begin{alignedat}{4}
G_F           &= 1.166379\times10^{-5}\;\mathrm{GeV}^{-2}, 
&\quad
m_W           &= 80.379\;\mathrm{GeV},      
&\quad
m_Z           &= 91.1876\;\mathrm{GeV},     \\[2pt]
v                  &= 246.22\;\mathrm{GeV},      
&\quad
\alpha   &= 1/132.184,                 
&\quad
s_{\theta}^2              &= 0.2230,                    \\[2pt]
c_{\theta}^2              &= 1 - s_{\theta}^2,                 
&\quad
m_t                &= 172.5\;\mathrm{GeV},       
&\quad
\Lambda            &= 1\;\mathrm{TeV},           \\[2pt]
\mu_R              &= m_Z,                  
&\quad
\Gamma_Z^{\rm SM}  &= 2.4941\;\mathrm{GeV},       
&\quad
\Gamma_{\rm had}^{\rm SM} &= 1.6944\;\mathrm{GeV},   \\[2pt]
R_b^{\rm SM}       &= 0.21582,                  
&\quad
A_b^{\rm SM}       &= 0.9347,                   
&\quad
A_{b,\mathrm{FB}}^{\rm SM} &= 0.1029\,.
\end{alignedat}
\end{equation}

\subsubsection*{Corrections to $\Gamma_{b}$}
Adopting the one‐loop result of Ref.~\cite{Haisch:2024wnw} and including the two‐loop contributions as described above, the relative shift in the \(Z\to b\bar b\) partial width, $\delta \Gamma_b$, reads 
\begin{equation}
\begin{aligned}
\nicefrac{\delta \Gamma_b^{\mathrm{1L}}}{\Gamma_Z^{\mathrm{SM}}}
&= (3.8320\,c_{QQ}^{1}
  +0.4065\,c_{QQ}^{8}
  -4.5839\,c_{Qt}^{1}) \times 10^{-4}
\\[6pt]
\nicefrac{\delta \Gamma_b^{\mathrm{1L+2L}}}{\Gamma_Z^{\mathrm{SM}}}
&=(3.8255\,c_{QQ}^{1}
  +0.4049\,c_{QQ}^{8}
  -4.5745\,c_{Qt}^{1}
  +0.0018\,c_{Qt}^{8}
  -0.0125\,c_{tt}^{1}) \times 10^{-4},
\end{aligned}
\label{eq:gammab_shift_app}
\end{equation}
where \(\delta\Gamma_b^{\mathrm{1L}}\) and \(\delta\Gamma_b^{\mathrm{1L+2L}}\) denote the one-loop and the combined one- and two-loop contributions, respectively.

\subsubsection*{Corrections to $R_c$, $R_\ell$, and $R_b$}
Implementing the above definitions and expanding to first order in $\delta\Gamma_b$, we obtain
\begin{equation}
    \begin{aligned}
\bar R_c &=
  \frac{\Gamma_c^{\rm SM}}
       {\Gamma_{\rm had}^{\rm SM}+\delta\Gamma_b}
  \;\simeq\;
  R_c^{\rm SM}\!\left(1-\frac{\delta\Gamma_b}{\Gamma_{\rm had}^{\rm SM}}\right),
&
\bar R_\ell &=
  \frac{\Gamma_{\rm had}^{\rm SM}+\delta\Gamma_b}
       {\Gamma_\ell^{\rm SM}}
  \;\simeq\;
  R_\ell^{\rm SM}\!\left(1+\frac{\delta\Gamma_b}{\Gamma_{\rm had}^{\rm SM}}\right),\notag
\end{aligned}
\end{equation}
\begin{equation}
\label{eq:Rb}
\bar R_b =
  \frac{\Gamma_b^{\rm SM}+\delta\Gamma_b}
       {\Gamma_{\rm had}^{\rm SM}+\delta\Gamma_b}
  \;\simeq\;
  R_b^{\rm SM}
  +\bigl(1-R_b^{\rm SM}\bigr)\,
   \frac{\delta\Gamma_b}{\Gamma_{\rm had}^{\rm SM}}\,.
\end{equation}
Replacing \(\delta\Gamma_b\to\delta\Gamma_b^{\rm 1L}\) or \(\delta\Gamma_b^{\rm 1L+2L}\) gives
\begin{equation}
\begin{aligned}
\nicefrac{\delta R_c^{\mathrm{1L}}}{R_c^{\rm SM}}
&= (-5.6404\,c_{QQ}^{1}
  -0.5983\,c_{QQ}^{8}
  +6.7472\,c_{Qt}^{1}) \times 10^{-4}
\\[6pt]
\nicefrac{\delta R_c^{\mathrm{1L+2L}}}{R_c^{\rm SM}}
&=(-5.6308\,c_{QQ}^{1}
  -0.5959\,c_{QQ}^{8}
  +6.7333\,c_{Qt}^{1}
  -0.0026\,c_{Qt}^{8}
  +0.0184\,c_{tt}^{1}) \times 10^{-4},
\\[6pt]
\nicefrac{\delta R_b^{\mathrm{1L}}}{R_b^{\rm SM}}
&= (20.494\,c_{QQ}^{1}
  +2.1742\,c_{QQ}^{8}
  -24.516\,c_{Qt}^{1}) \times 10^{-4}
\\[6pt]
\nicefrac{\delta R_b^{\mathrm{1L+2L}}}{R_b^{\rm SM}}
&=(20.459\,c_{QQ}^{1}
  +2.1654\,c_{QQ}^{8}
  -24.465\,c_{Qt}^{1}
  +0.0096\,c_{Qt}^{8}
  -0.0669\,c_{tt}^{1}) \times 10^{-4}.
\end{aligned}
\label{eq:gammab_shift_app}
\end{equation}

\subsubsection*{Corrections to $A_b$, $A_{b,\rm{FB}}$}
Using the one‐loop result of Ref.~\cite{Haisch:2024wnw}, the one‐loop shifts read
\begin{equation}
\begin{aligned}
\nicefrac{\delta A_{b}^{\mathrm{1L}}}{A_{b}^{\rm SM}}
&=(2.3648\,c_{QQ}^{1}
  +0.2508\,c_{QQ}^{8}
  -2.8288\,c_{Qt}^{1}) \times 10^{-4}
\\[6pt]
\nicefrac{\delta A_{b,\mathrm{FB}}^{\mathrm{1L}}}{A_{b,\mathrm{FB}}^{\rm SM}}
&= (2.3682\,c_{QQ}^{1}
  +0.2512\,c_{QQ}^{8}
  -2.8329\,c_{Qt}^{1}) \times 10^{-4}
\end{aligned} 
\end{equation}
where we have used the relation $A_{b,\rm{FB}}= \nicefrac{3}{4}A_{b}A_{e}$~\cite{ThomasArun:2023wbd}. Including the two‐loop shifts using the following relations~\cite{ThomasArun:2023wbd}:
\begin{equation}
\begin{aligned}
A_e \;=\; 2 \,\frac{g_V^\ell\,g_A^\ell}{(g_V^\ell)^2 + (g_A^\ell)^2}
\,,\qquad
A_f \;=\; 2 \,\frac{g_V^f\,g_A^f}{(g_V^f)^2 + (g_A^f)^2}
\,,
\end{aligned}
\end{equation}
we obtain the total corrections for the asymmetry observables:
\begin{equation}
\begin{aligned}
\nicefrac{\delta A_{b}^{\mathrm{1L+2L}}}{A_{b}^{\rm SM}}
&=(2.6316\,c_{QQ}^{1}
  +0.3447\,c_{QQ}^{8}
  -3.1499\,c_{Qt}^{1}
  +0.0062\,c_{Qt}^{8}
  +0.6133\,c_{tt}^{1}) \times 10^{-4},
\\[6pt]
\nicefrac{\delta A_{b,\mathrm{FB}}^{\mathrm{1L+2L}}}{A_{b,\mathrm{FB}}^{\rm SM}}
&= (24.593\,c_{QQ}^{1}
  +8.0737\,c_{QQ}^{8}
  -29.576\,c_{Qt}^{1}
  +0.5198\,c_{Qt}^{8}
  +51.083\,c_{tt}^{1}) \times 10^{-4}.
\end{aligned} 
\end{equation}

\section{Additional fit results}
We report here the 95\% CL bounds on each of the five WCs from each process, summarised in~\cref{tab:all-bounds}. The corresponding individual and marginalised limits from the combined fit are shown in~\cref{tab:combined_full}. All bounds are quoted at both linear and quadratic order in the EFT expansion.
\begin{table}[ht]
    \centering
    \renewcommand{\arraystretch}{1.0}
    \resizebox{\textwidth}{!}{%
    \input{tab_all_bounds}}
    \caption{95\% CL individual bounds from each process.}
    \label{tab:all-bounds}
\end{table}
\begin{table}[ht]
\centering
\begin{tabular}{l *{2}{r@{\,}c@{\,}l} *{2}{r@{\,}c@{\,}l}}
\toprule
 & \multicolumn{6}{c}{Individual} 
 & \multicolumn{6}{c}{Marginalised} \\
\cmidrule(lr){2-7} \cmidrule(lr){8-13}
 & \multicolumn{3}{c}{$\mathcal O(\Lambda^{-2})$}
 & \multicolumn{3}{c}{$\mathcal O(\Lambda^{-4})$}
 & \multicolumn{3}{c}{$\mathcal O(\Lambda^{-2})$}
 & \multicolumn{3}{c}{$\mathcal O(\Lambda^{-4})$} \\
\cmidrule(lr){2-4} \cmidrule(lr){5-7} \cmidrule(lr){8-10} \cmidrule(lr){11-13}
$c^1_{tt}$
 & [ & -9.12,\, -0.89 & ]
 & [ & -1.66,\,  1.49 & ]
 & [ & -16.30,\, -1.45 & ]
 & [ & -1.50,\,  1.52 & ] \\
$c^1_{QQ}$
 & [ & -0.99,\,  2.41 & ]
 & [ & -0.84,\,  3.03 & ]
 & [ & -8.42,\,  8.50 & ]
 & [ & -0.97,\,  5.94 & ] \\
$c^8_{QQ}$
 & [ & -11.39,\, 11.46 & ]
 & [ & -8.60,\,  10.84& ]
 & [ & -75.66,\, 48.77& ]
 & [ & -19.43,\, 5.85 & ] \\
$c^1_{Qt}$
 & [ & -1.74,\,  1.03 & ]
 & [ & -2.30,\,  1.17 & ]
 & [ & -10.99,\,  6.71 & ]
 & [ & -1.37,\,  2.69 & ] \\
$c^8_{Qt}$
 & [ & -7.93,\, 12.88 & ]
 & [ & -4.39,\,  6.63 & ]
 & [ & -9.50,\, 58.36 & ]
 & [ & -4.09,\,  6.20 & ] \\
\bottomrule
\end{tabular}
\caption{95\% CL individual and marginalised bounds from the combined fit.}
\label{tab:combined_full}
\end{table}

\bibliographystyle{JHEP}
\bibliography{bibliography}

\end{document}

%% file: Figures/Fig2.1.tex
\begin{tikzpicture} 
            \begin{feynman}[small]
                \vertex  (g1)  {$g$};
                \vertex  (gtt1) [dot,scale=\sizedot,right = of g1] {};
                \vertex (4F) [square dot, scale=\sizesqdot,below right= of gtt1,color=black] {};
                \vertex  (gtt2) [dot,scale=\sizedot,below left = of 4F] {};
                \vertex  (g2) [left=of gtt2]  {$g$};
                \vertex (htt) [dot,scale=\sizedot,right = 20 pt  of 4F] {};
                \vertex (h) [right = of htt] {$H$};

                \diagram* {
                    (g1)  -- [gluon] (gtt1),
                    (g2) -- [gluon] (gtt2),
                    (h)  -- [scalar] (htt),
                    (gtt1) -- [fermion] (4F) 
                    -- [fermion] (gtt2)
                     -- [fermion] (gtt1), 
                     (4F) -- [fermion, half left] (htt) -- [fermion, half left] (4F)
                };
            \end{feynman}
        \end{tikzpicture}

%% file: Figures/Fig2.2.tex
\begin{tikzpicture} 
            \begin{feynman}[small]
                \vertex  (g1)  {$g$};
                \vertex  (gtt1) [dot,scale=\sizedot,right = of g1] {};
                \vertex (htt) [dot,scale=\sizedot,below right = of gtt1] {};
                \vertex (4F) [square dot, scale = \sizesqdot,color=black,below right= 14 pt of gtt1] {};
                \vertex (inv1) [scale=0.01,above right = 18 pt of 4F] {};
                \vertex  (gtt2) [dot,scale=\sizedot,below left = of htt] {};
                \vertex  (g2) [left=of gtt2]  {$g$};
                \vertex (h) [right = of htt] {$H$};
                \diagram* {
                    (g1)  -- [gluon] (gtt1),
                    (g2) -- [gluon] (gtt2),
                    (h)  -- [scalar] (htt),
                    (gtt1) -- [fermion] (4F) -- [fermion] (htt) 
                    -- [fermion] (gtt2)
                     -- [fermion] (gtt1), 
                     (4F) -- [fermion, half left] (inv1) -- [fermion, half left] (4F)
};
            \end{feynman}
        \end{tikzpicture}

%% file: Figures/Fig2.3.tex
\begin{tikzpicture} 
            \begin{feynman}[small]
                \vertex  (g1)  {$g$};
                 \vertex (gtt1) [dot,scale=\sizedot,right= 28 pt of g1] {};
                \vertex  (4F) [square dot, scale=\sizesqdot, right = 20 pt of gtt1,color=black] {};
                \vertex (htt) [dot,scale=\sizedot,below right = of 4F] {};
                \vertex  (gtt2) [dot,scale=\sizedot,below left = of htt] {};
                \vertex  (g2) [left= 50 pt of gtt2]  {$g$};
                \vertex (h) [right = of htt] {$H$};

                \diagram* {
                    (g1)  -- [gluon] (gtt1),
                    (g2) -- [gluon] (gtt2),
                    (h)  -- [scalar] (htt),
                    (gtt1) -- [fermion, half right] (4F) 
                    -- [fermion,half right] (gtt1),
                     (4F) -- [fermion] (htt) -- [fermion] (gtt2) -- [fermion] (4F)
                };
            \end{feynman}
        \end{tikzpicture}

%% file: table_pred_pTH.tex
    \begin{tabularx}{0.87\textwidth}{l*{6}{>{\centering\arraybackslash}X}}
\hline\\[-0.4cm]
    &\multicolumn{6}{c}{$d\sigma_{\rm NLO}/{dp_T^H} \rm [pb]$}\\[0.1cm]
\hline\\[-0.4cm]
    & \tiny $p_T^H < 60 \rm \, GeV$ 
    & \tiny $ 60 \leq p_T^H < 120 \rm \, GeV$ 
    & \tiny $ 120 \leq p_T^H < 200 \rm \, GeV$ 
    & \tiny $ 200 \leq p_T^H < 300 \rm \, GeV$ 
    & \tiny $ 300 \leq p_T^H < 450 \rm \, GeV$ 
    & \tiny  $ p_T^H \geq 450 \rm \, GeV$  
\\[0.1cm]
\hline\\[-0.4cm]
 SM 
 & \tiny $\expnumber{1.197}{-1}\pm 0.054\% ^{+6\%}_{-9.08\%}$
 & \tiny $\expnumber{1.785}{-1}\pm 0.052\% ^{+5.94\%}_{-9.17\%}$
 & \tiny $\expnumber{1.258}{-1}\pm 0.065\% ^{+5.97\%}_{-9.43\%}$
 & \tiny $\expnumber{5.203}{-2}\pm 0.096\% ^{+5.47\%}_{-9.61\%}$
 & \tiny $\expnumber{1.888}{-2}\pm 0.141\% ^{+3.36\%}_{-9.24\%}$
 & \tiny $\expnumber{5.179}{-3}\pm 0.28\% ^{+1.76\%}_{-6.8\%}$ \\
\hline\\[-0.4cm]
 & \multicolumn{6}{c}{$\mathcal{O}(c_{i}/\Lambda^{2})$} \\[0.1cm]
 $c_{tt}^{1}$ 
 & \tiny $\expnumber{-3.693}{-5}\pm 1.878\% ^{+55.11\%}_{-34.6\%}$
 & \tiny $\expnumber{-1.173}{-4}\pm 0.867\% ^{+49.39\%}_{-31.81\%}$
 & \tiny $\expnumber{-1.223}{-4}\pm 1.11\% ^{+56.16\%}_{-35.46\%}$
 & \tiny $\expnumber{1.134}{-6}\pm 139\% ^{+1632\%}_{-2106\%}$
 & \tiny $\expnumber{9.791}{-5}\pm 1.872\% ^{+21.88\%}_{-18.05\%}$
 & \tiny $\expnumber{1.567}{-4}\pm 1.387\% ^{+31.86\%}_{-22.53\%}$ \\
 $c_{QQ}^{1}$ 
 & \tiny $\expnumber{-9.997}{-5}\pm 1.227\% ^{+36.9\%}_{-25.25\%}$
 & \tiny $\expnumber{-1.911}{-4}\pm 0.96\% ^{+38.41\%}_{-26.06\%}$
 & \tiny $\expnumber{-1.63}{-4}\pm 1.896\% ^{+42.95\%}_{-28.45\%}$
 & \tiny $\expnumber{-4.519}{-5}\pm 6.025\% ^{+64.81\%}_{-39.8\%}$
 & \tiny $\expnumber{3.056}{-5}\pm 9.49\% ^{+32.56\%}_{-28.18\%}$
 & \tiny $\expnumber{7.125}{-5}\pm 5.559\% ^{+31.13\%}_{-22.19\%}$ \\
 $c_{QQ}^{8}$ 
 & \tiny $\expnumber{2.735}{-5}\pm 3.993\% ^{+11.41\%}_{-14.46\%}$
 & \tiny $\expnumber{5.53}{-5}\pm 3.214\% ^{+9.25\%}_{-13.25\%}$
 & \tiny $\expnumber{8.329}{-5}\pm 2.527\% ^{+13.33\%}_{-13.98\%}$
 & \tiny $\expnumber{8.667}{-5}\pm 2.701\% ^{+15.39\%}_{-12.49\%}$
 & \tiny $\expnumber{8.205}{-5}\pm 3.134\% ^{+24.33\%}_{-17.51\%}$
 & \tiny $\expnumber{7.729}{-5}\pm 4.074\% ^{+32.33\%}_{-22.3\%}$ \\
 $c_{Qt}^{1}$ 
 & \tiny $\expnumber{-1.262}{-3}\pm 0.246\% ^{+31.97\%}_{-22.67\%}$
 & \tiny $\expnumber{-1.93}{-3}\pm 0.203\% ^{+32.61\%}_{-22.98\%}$
 & \tiny $\expnumber{-1.338}{-3}\pm 0.258\% ^{+33.74\%}_{-23.52\%}$
 & \tiny $\expnumber{-4.837}{-4}\pm 0.553\% ^{+34.96\%}_{-24.05\%}$
 & \tiny $\expnumber{-1.277}{-4}\pm 1.555\% ^{+34.69\%}_{-23.78\%}$
 & \tiny $\expnumber{-3.17}{-5}\pm 6.399\% ^{+31.59\%}_{-22.27\%}$ \\
 $c_{Qt}^{8}$ 
 & \tiny $\expnumber{-1.149}{-4}\pm 1.008\% ^{+39.12\%}_{-28.69\%}$
 & \tiny $\expnumber{-1.266}{-4}\pm 1.521\% ^{+45.7\%}_{-33.33\%}$
 & \tiny $\expnumber{3.364}{-6}\pm 74.484\% ^{+389\%}_{-535\%}$
 & \tiny $\expnumber{9.672}{-5}\pm 2.479\% ^{+20.53\%}_{-14.5\%}$
 & \tiny $\expnumber{1.129}{-4}\pm 2.405\% ^{+28.3\%}_{-19.84\%}$
 & \tiny $\expnumber{1.088}{-4}\pm 3.605\% ^{+33.57\%}_{-23.02\%}$ \\
\hline\\[-0.4cm]
 & \multicolumn{6}{c}{$\mathcal{O}(c_{i}^2/\Lambda^{4})$} \\[0.1cm]
 $c_{tt}^{1}$ 
 & $\times$ 
 & $\times$
 & $\times$ 
 & $\times$ 
 & $\times$ 
 & $\times$ \\
 $c_{QQ}^{1}$ 
 & \tiny $\expnumber{2.89}{-5}\pm 2.048\% ^{+4.14\%}_{-5.08\%}$
 & \tiny $\expnumber{5.913}{-5}\pm 1.867\% ^{+4.61\%}_{-4.27\%}$
 & \tiny $\expnumber{6.764}{-5}\pm 1.804\% ^{+4.94\%}_{-3.84\%}$
 & \tiny $\expnumber{5.253}{-5}\pm 2.259\% ^{+6.82\%}_{-4.48\%}$
 & \tiny $\expnumber{3.687}{-5}\pm 3.338\% ^{+8.42\%}_{-5.56\%}$
 & \tiny $\expnumber{2.779}{-5}\pm 4.385\% ^{+8.37\%}_{-5.23\%}$ \\
 $c_{QQ}^{8}$ 
 & \tiny $\expnumber{4.415}{-6}\pm 3.202\% ^{+8.53\%}_{-5.52\%}$
 & \tiny $\expnumber{8.902}{-6}\pm 2.504\% ^{+7.99\%}_{-4.62\%}$
 & \tiny $\expnumber{1.034}{-5}\pm 2.034\% ^{+9.56\%}_{-5.3\%}$
 & \tiny $\expnumber{8.285}{-6}\pm 2.89\% ^{+10.59\%}_{-5.96\%}$
 & \tiny $\expnumber{5.831}{-6}\pm 4.934\% ^{+12\%}_{-6.83\%}$
 & \tiny $\expnumber{4.937}{-6}\pm 6.78\% ^{+14.48\%}_{-8.38\%}$ \\
 $c_{Qt}^{1}$ 
 & \tiny $\expnumber{2.833}{-5}\pm 1.938\% ^{+4.06\%}_{-5.02\%}$
 & \tiny $\expnumber{6.055}{-5}\pm 1.45\% ^{+4.17\%}_{-5.06\%}$
 & \tiny $\expnumber{6.545}{-5}\pm 1.927\% ^{+5.11\%}_{-3.51\%}$
 & \tiny $\expnumber{5.113}{-5}\pm 2.355\% ^{+6.15\%}_{-3.97\%}$
 & \tiny $\expnumber{3.639}{-5}\pm 3.024\% ^{+7.48\%}_{-4.79\%}$
 & \tiny $\expnumber{3.042}{-5}\pm 4.135\% ^{+12.52\%}_{-8.65\%}$ \\
 $c_{Qt}^{8}$ 
 & \tiny $\expnumber{4.303}{-6}\pm 3.461\% ^{+8.26\%}_{-5.78\%}$
 & \tiny $\expnumber{8.2}{-6}\pm 2.807\% ^{+8.72\%}_{-6.28\%}$
 & \tiny $\expnumber{1.01}{-5}\pm 2.563\% ^{+9.22\%}_{-5.09\%}$
 & \tiny $\expnumber{7.497}{-6}\pm 7.655\% ^{+11.93\%}_{-7.69\%}$
 & \tiny $\expnumber{4.598}{-6}\pm 11.933\% ^{+15.1\%}_{-13.59\%}$
 & \tiny $\expnumber{4.561}{-6}\pm 8.879\% ^{+12.56\%}_{-6.43\%}$ \\
 \hline\\[-0.4cm]
 & \multicolumn{6}{c}{$\mathcal{O}(c_{i}c_{j}/\Lambda^{4})$} \\[0.1cm]
 $c_{QQ}^{1}c_{Qt}^{1}$ 
 & \tiny $\expnumber{1.49}{-5}\pm 8.241\% ^{+8.56\%}_{-6.91\%}$
 & \tiny $\expnumber{2.817}{-5}\pm 7.635\% ^{+9.7\%}_{-8.02\%}$
 & \tiny $\expnumber{2.928}{-5}\pm 9.17\% ^{+10.63\%}_{-6.63\%}$
 & \tiny $\expnumber{1.762}{-5}\pm 14.089\% ^{+10.58\%}_{-6.63\%}$
 & \tiny $\expnumber{1.086}{-5}\pm 21.595\% ^{+8.73\%}_{-8.76\%}$
 & \tiny $\expnumber{-2.153}{-6}\pm 118\% ^{+62.83\%}_{-68.99\%}$ \\
 $c_{QQ}^{8}c_{Qt}^{8}$
 & \tiny $\expnumber{3.253}{-6}\pm 11.559\% ^{+2.76\%}_{-5.77\%}$
 & \tiny $\expnumber{5.878}{-6}\pm 10.101\% ^{+10.34\%}_{-13.75\%}$
 & \tiny $\expnumber{7.997}{-6}\pm 7.513\% ^{+4.73\%}_{-6.71\%}$
 & \tiny $\expnumber{4.398}{-6}\pm 19.689\% ^{+9.78\%}_{-12.84\%}$
 & \tiny $\expnumber{4.704}{-6}\pm 18.524\% ^{+5.68\%}_{-1.91\%}$
 & \tiny $\expnumber{6.259}{-7}\pm 123\% ^{+109\%}_{-122\%}$ \\
\hline
\end{tabularx}

%% file: table_pred_incl_all.tex
\begin{tabularx}{1.0\textwidth}{l*{7}{>{\centering\arraybackslash}X}}
\hline\\[-0.4cm]
    &\multicolumn{7}{c}{$\sigma^{\rm proc.}_{\rm (N)LO} [\rm pb] \rm$}\\[0.1cm]
\hline\\[-0.4cm]
    & \tiny $ \sigma^{t\bar{t}}_{\rm LO}$
    & \tiny $ \sigma^{t\bar{t}}_{\rm NLO}$
    & \tiny $ K_{t\bar{t}}$
    & \tiny $ \sigma^{t\bar{t}H}_{\rm LO}$
    & \tiny $ \sigma^{t\bar{t}H}_{\rm NLO}$
    & \tiny $ K_{t\bar{t}H}$ 
    & \tiny $ \sigma^{t\bar{t}t\bar{t}}_{\rm LO}$
\\[0.1cm]
\hline\\[-0.4cm]
 SM 
 & \tiny $\expnumber{5.028}{+2}\pm 0.003\% ^{+29.67\%}_{-21.44\%}$
 & \tiny $\expnumber{7.532}{+2}\pm 0.004\% ^{+11.75\%}_{-11.9\%}$ 
 & \tiny 1.498 
 & \tiny $\expnumber{4.005}{-1}\pm 0.005\% ^{+31.45\%}_{-22.25\%}$
 & \tiny $\expnumber{5.003}{-1}\pm 0.009\% ^{+5.71\%}_{-9.24\%}$ 
 & \tiny 1.249 
 & \tiny $\expnumber{6.754}{-3}\pm 0.016\% ^{+62.66\%}_{-35.48\%}$\\ 
 \hline\\[-0.4cm]
 & \multicolumn{7}{c}{$\mathcal{O}(c_{i}/\Lambda^{2})$} \\[0.1cm]
 $c_{tt}^{1}$
 & $\times$
 & \tiny $\expnumber{2.376}{-1}\pm 0.461\% ^{+24.03\%}_{-18.41\%}$
 & $\times$ 
 & $\times$
 & \tiny $\expnumber{-2.089}{-5}\pm 17.867\% ^{+494\%}_{-339\%}$
 & $\times$ 
 & \tiny $\expnumber{-1.185}{-3}\pm 0.026\% ^{+36.07\%}_{-24.93\%}$\\
 $c_{QQ}^{1}$ 
 & $\times$ 
 & \tiny $\expnumber{-3.323}{-2}\pm 1.731\% ^{+53.94\%}_{-34.92\%}$
 & $\times$ 
 & $\times$ 
 & \tiny $\expnumber{-3.975}{-4}\pm 1.688\% ^{+46.88\%}_{-30.48\%}$
 & $\times$ 
 & \tiny $\expnumber{-7.053}{-4}\pm 0.024\% ^{+37.35\%}_{-25.48\%}$\\
 $c_{QQ}^{8}$ 
 & \tiny $\expnumber{5.961}{-2}\pm 0.123\% ^{+27.97\%}_{-25.65\%}$
 & \tiny $\expnumber{1.209}{-1}\pm 0.406\% ^{+10.94\%}_{-12.13\%}$
 & \tiny 2.028 
 & \tiny $\expnumber{1.708}{-4}\pm 0.393\% ^{+22.3\%}_{-21.22\%}$
 & \tiny $\expnumber{4.119}{-4}\pm 1.3\% ^{+12.4\%}_{-11.31\%}$
 & \tiny 2.411 
 & \tiny $\expnumber{-2.351}{-4}\pm 0.027\% ^{+37.39\%}_{-25.50\%}$\\
 $c_{Qt}^{1}$ 
 & $\times$
 & \tiny $-2.538\pm 0.053\% ^{+29.22\%}_{-21.29\%}$
 & $\times$ 
 & $\times$
 & \tiny $\expnumber{-5.175}{-3}\pm 0.138\% ^{+33.01\%}_{-23.16\%}$
 & $\times$ 
 & \tiny $\expnumber{7.191}{-4}\pm 0.024\% ^{+41.42\%}_{-27.19\%}$\\
 $c_{Qt}^{8}$ 
 & \tiny $\expnumber{5.951}{-2}\pm 0.135\% ^{+27.97\%}_{-25.65\%}$
 & \tiny $\expnumber{-1.116}{-1}\pm 0.48\% ^{+37.42\%}_{-30.59\%}$
 & \tiny 1.875
 & \tiny $\expnumber{1.702}{-4}\pm 0.464\% ^{+22.28\%}_{-21.21\%}$
 & \tiny $\expnumber{8.029}{-5}\pm 7.662\% ^{+33.34\%}_{-40.45\%}$ 
 & \tiny 0.471  
 & \tiny $\expnumber{-3.026}{-4}\pm 0.029\% ^{+29.98\%}_{-22.35\%}$\\
\hline\\[-0.4cm]
 & \multicolumn{7}{c}{$\mathcal{O}(c_{i}^2/\Lambda^{4})$} \\[0.1cm]
 $c_{tt}^{1}$ 
 & $\times$
 & $\times$
 & $\times$ 
 & $\times$
 & $\times$
 & $\times$ 
 & \tiny $\expnumber{4.342}{-3}\pm 0.021\% ^{+46.06\%}_{-29.23\%}$\\
 $c_{QQ}^{1}$ 
 & \tiny $\expnumber{2.739}{-2}\pm 0.233\% ^{+13.52\%}_{-16.73\%}$
 & \tiny $\expnumber{5.863}{-2}\pm 0.378\% ^{+6.51\%}_{-7.68\%}$
 & \tiny 2.14
 & \tiny $\expnumber{1.608}{-4}\pm 0.686\% ^{+9.34\%}_{-12.45\%}$
 & \tiny $\expnumber{2.728}{-4}\pm 0.84\% ^{+5.9\%}_{-3.77\%}$
 & \tiny 1.696 
 & \tiny $\expnumber{1.085}{-3}\pm 0.029\% ^{+46\%}_{-29.19\%}$\\
 $c_{QQ}^{8}$ 
 & \tiny $\expnumber{6.061}{-3}\pm 0.3\% ^{+13.53\%}_{-16.74\%}$
 & \tiny $\expnumber{1.029}{-2}\pm 0.365\% ^{+6.82\%}_{-4.65\%}$
 & \tiny 1.697 
 & \tiny $\expnumber{3.579}{-5}\pm 0.826\% ^{+9.33\%}_{-12.44\%}$
 & \tiny $\expnumber{4.271}{-5}\pm 1.275\% ^{+10.23\%}_{-5.71\%}$
 & \tiny 1.193 
 & \tiny $\expnumber{1.208}{-4}\pm 0.045\% ^{+45.98\%}_{-29.18\%}$\\
 $c_{Qt}^{1}$ 
 & \tiny $\expnumber{2.73}{-2}\pm 0.266\% ^{+13.54\%}_{-16.75\%}$
 & \tiny $\expnumber{5.877}{-2}\pm 0.378\% ^{+6.68\%}_{-7.78\%}$
 & \tiny 2.152
 & \tiny $\expnumber{1.616}{-4}\pm 0.614\% ^{+9.31\%}_{-12.42\%}$
 & \tiny $\expnumber{2.722}{-4}\pm 0.789\% ^{+5.96\%}_{-3.8\%}$
 & \tiny 1.684 
 & \tiny $\expnumber{1.47}{-3}\pm 0.048\% ^{+46.21\%}_{-29.28\%}$\\
 $c_{Qt}^{8}$ 
 & \tiny $\expnumber{6.047}{-3}\pm 0.32\% ^{+13.53\%}_{-16.74\%}$
 & \tiny $\expnumber{9.186}{-3}\pm 0.77\% ^{+9.22\%}_{-6.16\%}$
 & \tiny 1.519
 & \tiny $\expnumber{3.597}{-5}\pm 0.982\% ^{+9.29\%}_{-12.41\%}$
 & \tiny $\expnumber{3.926}{-5}\pm 2.326\% ^{+10.61\%}_{-6.52\%}$
 & \tiny 1.091 
 & \tiny $\expnumber{3.545}{-4}\pm 0.055\% ^{+45.89\%}_{-29.15\%}$\\
 \hline\\[-0.4cm]
 & \multicolumn{7}{c}{$\mathcal{O}(c_{i}c_{j}/\Lambda^{4})$} \\[0.1cm]
 $c_{QQ}^{1}c_{Qt}^{1}$ 
 & \tiny $\expnumber{1.41}{-2}\pm 0.778\% ^{+15.62\%}_{-18.47\%}$
 & \tiny $\expnumber{1.128}{-2}\pm 3.178\% ^{+11.8\%}_{-10.75\%}$
 & \tiny 0.8
 & \tiny $\expnumber{1.585}{-4}\pm 1.231\% ^{+10.1\%}_{-13.13\%}$
 & \tiny $\expnumber{9.87}{-5}\pm 4.431\% ^{+11.21\%}_{-7.92\%}$
 & \tiny 0.622
 & \tiny $\expnumber{-3.51}{-4}\pm 0.261\% ^{+43.39\%}_{-28.12\%}$\\
 $c_{QQ}^{1}c_{tt}^{1}$
 & $-$ 
 & $-$ 
 & $\times$ 
 & $-$ 
 & $-$ 
 & $\times$ 
 & \tiny $\expnumber{3.566}{-4}\pm 0.365\% ^{+41.6\%}_{-27.33\%}$\\
 $c_{QQ}^{1}c_{QQ}^{8}$ 
 & $-$ 
 & \tiny $\expnumber{1.066}{-3}\pm 23.367\% ^{+24.19\%}_{-23.49\%}$
 & $\times$ 
 & $-$
 & $-$
 & $\times$ 
 & \tiny $\expnumber{7.241}{-4}\pm 0.072\% ^{+46.07\%}_{-29.23\%}$\\
 $c_{QQ}^{1}c_{Qt}^{8}$ 
 & $-$ 
 & $-$
 & $\times$ 
 & $-$ 
 & $-$
 & $\times$  
 & \tiny $\expnumber{1.314}{-4}\pm 0.381\% ^{+42.77\%}_{-27.84\%}$\\
 $c_{Qt}^{1}c_{tt}^{1}$ 
 & $-$
 & \tiny $\expnumber{5.545}{-4}\pm 45.484\% ^{+14.2\%}_{-16.01\%}$
 & $\times$ 
 & $-$
 & $-$
 & $\times$
 & \tiny $\expnumber{-7.075}{-4}\pm 0.202\% ^{+43.31\%}_{-28.07\%}$\\
 $c_{Qt}^{1}c_{QQ}^{8}$ 
 & $-$
 & $-$
 & $\times$ 
 & $-$
 & $-$
 & $\times$  
 & \tiny $\expnumber{-1.18}{-4}\pm 0.723\% ^{+43.31\%}_{-28.06\%}$\\
 $c_{Qt}^{1}c_{Qt}^{8}$ 
 & $-$
 & \tiny $\expnumber{-2.956}{-3}\pm 9.905\% ^{+28.22\%}_{-26.3\%}$
 & $\times$
 & $-$
 & \tiny $\expnumber{-1.689}{-5}\pm 19.597\% ^{+23.96\%}_{-23.82\%}$
 & $\times$ 
 & \tiny $\expnumber{-8.382}{-5}\pm 1.078\% ^{+41.67\%}_{-27.38\%}$\\
 $c_{tt}^{1}c_{QQ}^{8}$ 
 & $-$
 & $-$
 & $\times$ 
 & $-$
 & \tiny $\expnumber{1.931}{-6}\pm 37.788\% ^{+26.67\%}_{-25.98\%}$
 & $\times$ 
 & \tiny $\expnumber{1.181}{-4}\pm 1.002\% ^{+41.56\%}_{-27.31\%}$\\
 $c_{tt}^{1}c_{Qt}^{8}$ 
 & $-$
 & \tiny $\expnumber{4.615}{-4}\pm 16.652\% ^{+30.32\%}_{-27.57\%}$
 & $\times$ 
 & $-$
 & $-$
 & $\times$
 & \tiny $\expnumber{2.603}{-4}\pm 0.461\% ^{+42.62\%}_{-27.77\%}$\\
 $c_{QQ}^{8}c_{Qt}^{8}$ 
 & \tiny $\expnumber{3.177}{-3}\pm 0.965\% ^{+15.56\%}_{-18.42\%}$
 & \tiny $\expnumber{3.907}{-3}\pm 2.285\%  ^{+2.25\%}_{-3.99\%}$
 & \tiny 1.229
 & \tiny $\expnumber{3.45}{-5}\pm 2.034\% ^{+10.22\%}_{-13.24\%}$
 & \tiny $\expnumber{2.685}{-5}\pm 5.113\% ^{+7.94\%}_{-10.45\%}$
 & \tiny 0.778 
 & \tiny $\expnumber{4.391}{-5}\pm 0.584\% ^{+42.65\%}_{-27.77\%}$\\
\hline
\end{tabularx}

%% file: tab_all_bounds.tex
\centering
\begin{tabular}{l c *{6}{r@{\,}c@{\,}l}}
\toprule
 & Order 
  & \multicolumn{3}{c}{$t\bar tH$}
  & \multicolumn{3}{c}{$t\bar{t}t\bar t$}
  & \multicolumn{3}{c}{$t\bar t b\bar b$}
  & \multicolumn{3}{c}{$gg\!\to\!H$}
  & \multicolumn{3}{c}{$t\bar t$}
  & \multicolumn{3}{c}{EWPO} \\
\cmidrule(lr){3-5}\cmidrule(lr){6-8}\cmidrule(lr){9-11}%
\cmidrule(lr){12-14}\cmidrule(lr){15-17}\cmidrule(lr){18-20}
\multirow{2}{*}{$c^1_{tt}$}
 & $\mathcal{O}(\Lambda^{-2})$
    & [ & -79.98,\,37.36 & ]
    & [ & -11.31,\,1.42  & ]
    & \multicolumn{3}{c}{–}
    & \multicolumn{3}{c}{–}
    & [ & -6.94,\,11.72  & ]
     & [ & -14.92,\,-1.68 & ] \\
 & $\mathcal{O}(\Lambda^{-4})$
    & [ & -79.98,\,37.36 & ]
    & [ & -1.62,\,1.89   & ]
    & \multicolumn{3}{c}{–}
    & \multicolumn{3}{c}{–}
    & [ & -6.94,\,11.72  & ]
   & [ & -14.92,\,-1.68 & ] \\

\addlinespace
\multirow{2}{*}{$c^1_{QQ}$}
  & $\mathcal{O}(\Lambda^{-2})$
     & \multicolumn{3}{c}{–}
    & [ & -19.00,\,2.40  & ]
    & \multicolumn{3}{c}{–}
    & \multicolumn{3}{c}{–}
    & [ & -14.75,\,22.89 & ]
     & [ & -0.94,\,2.49   & ] \\
 & $\mathcal{O}(\Lambda^{-4})$
    & [ & -13.20,\,11.92 & ]
    & [ & -3.20,\,3.85   & ]
    & [ & -9.31,\,9.32   & ]
    & \multicolumn{3}{c}{–}
    & [ & -14.06,\,9.36  & ]
    & [ & -0.94,\,2.49   & ] \\

\addlinespace
\multirow{2}{*}{$c^8_{QQ}$}
   & $\mathcal{O}(\Lambda^{-2})$
    & \multicolumn{3}{c}{–}
    & [ & -57.01,\,7.20  & ]
    & [ & -39.53,\,91.23 & ]
    & \multicolumn{3}{c}{–}
    & [ & -16.18,\,26.89 & ]
   & [ & -13.31,\,16.79& ] \\
  & $\mathcal{O}(\Lambda^{-4})$
    & [ & -38.30,\,24.08 & ]
    & [ & -9.60,\,11.55  & ]
    & [ & -21.91,\,17.67 & ]
    & \multicolumn{3}{c}{–}
    & [ & -41.86,\,17.41 & ]
     & [ & -13.31,\,16.79& ] \\

\addlinespace
\multirow{2}{*}{$c^1_{Qt}$}
  & $\mathcal{O}(\Lambda^{-2})$
    & [ & -26.19,\,68.73 & ]
    & [ & -2.35,\,18.64  & ]
    & \multicolumn{3}{c}{–}
    & [ & -3.14,\,11.23  & ]
    & [ & -6.77,\,9.34   & ]
    & [ & -2.08,\,0.78   & ]\\
  & $\mathcal{O}(\Lambda^{-4})$
    & [ & -10.40,\,13.92 & ]
    & [ & -3.27,\,2.78   & ]
    & [ & -9.39,\,9.26   & ]
    & [ & -3.16,\,10.93  & ]
    & [ & -7.27,\,9.69   & ]
    & [ & -2.08,\,0.78   & ] \\

\addlinespace
\multirow{2}{*}{$c^8_{Qt}$}
  & $\mathcal{O}(\Lambda^{-2})$
    & [ & -117.00,\,39.40& ]
    & [ & -44.30,\,5.59  & ]
    & [ & -38.41,\,88.66 & ]
    & [ & -6.71,\,42.30  & ]
     & [ & -8.69,\,18.14  & ]
    & \multicolumn{3}{c}{–} \\
   & $\mathcal{O}(\Lambda^{-4})$
    & [ & -44.52,\,22.92 & ]
    & [ & -5.73,\,6.59   & ]
    & [ & -21.56,\,17.34 & ]
    & [ & -7.06,\,39.16  & ]
      & [ & -56.11,\,19.70  & ]
    & \multicolumn{3}{c}{–} \\
\bottomrule
\end{tabular}